\documentclass[preprint,authoryear,12pt]{elsarticle}

\usepackage{amssymb}

\newcommand{\PL}{$p(\lambda)$}
\newcommand{\sis}{$\sigma$}
\newcommand{\mbh}{$M_{\rm bh}$}
\newcommand{\msun}{$M_{\odot}$}

\newcommand{\lam}{$\lambda$}
\newcommand{\mstar}{$M_{\rm star}$}
\newcommand{\mbulge}{$M_{\rm bulge}$}

\newcommand{\ergs}{${\rm erg\, s^{-1}}$}

\newcommand{\re}{${r_e}$}

\begin{document}

\begin{frontmatter}


\title{The Demography of Super-Massive Black Holes:
Growing Monsters at the Heart of Galaxies}
\author{Francesco Shankar}
\ead{shankar@mpa-garching.mpg.de}
\address{Max-Planck-Institut f\"{u}r Astrophysik,
Karl-Schwarzschild-Str. 1, D-85748, Garching, Germany}

\begin{abstract}
Supermassive black holes (BHs) appear to be ubiquitous at the center of
all galaxies which have been observed at high enough sensitivities and
resolution with the Hubble Space Telescope. Their masses are found to be
tightly linked with the masses and velocity dispersions of their host
galaxies. On the other hand,  BHs are widely held to constitute the
central engines of quasars and active galactic nuclei (AGN) in general.
It is however still unclear how BHs have grown, and whether they have
co-evolved with their hosts.
In this Review I discuss how, in ways independent of specific models,
constraints on the growth history of BHs and their host galaxies have
been set by matching the statistics of local BHs to the emissivity,
number density, and clustering properties of AGNs at different
cosmological epochs.
I also present some new results obtained through a novel numerical
code which evolves the BH mass function and clustering adopting broad
distributions of Eddington ratios.
I finally review BH evolution in a wider cosmological context,
connecting BH growth to galaxy evolution.
\end{abstract}


\end{frontmatter}

\section{Introduction}
\label{intro}

Supermassive black holes (BHs) are common at the center of most, if
not all, galaxies (e.g., Magorrian et al., 1998; Ferrarese and Merritt, 2000; Gebhardt et al., 2000). As it will
be discussed below, their
mass is tightly linked with several properties of their host galaxies.
However, it still constitutes a major challenge understanding how these BHs
have grown, and if they followed the mass assembly history of their
host galaxies.

The discovery of such strong correlations lent further support to the seminal
idea, already proposed more than forty years back \citep[e.g.,][]{Salpeter64,Lynden69},
that quasars and, more generally, active galactic nuclei (AGNs) are powered by mass accretion onto central BHs, which should now lie at the center
of galaxies as relics of a past luminous phase.
The only process in fact able to generate
the high and highly concentrated powers observed in quasars is the release of
gravitational energy from an infalling body of mass $m$ onto an extremely compact
object of mass $M$.
This process can then generate energies up to a maximum of
$E=GMm/5R_g=0.1\, m c^2$, with the Schwarzschild radius $R_g=2GM/c^2\sim 3\times 10^{13} (M/10^8\, {\rm M_{\odot}})$ cm. Gravitational
processes can therefore release significant fractions of the rest-mass energy, orders
of magnitude more efficiently than the energy release from stellar nuclear reactions.
For example, a quasar with luminosity of $L=10^{46}\, $\ergs\ can be easily generated by
a massive object of $10^8$\msun\ confined within a region of size $\le 10^{15}\, $cm.

\begin{figure}
\includegraphics[angle=00,scale=1.3]{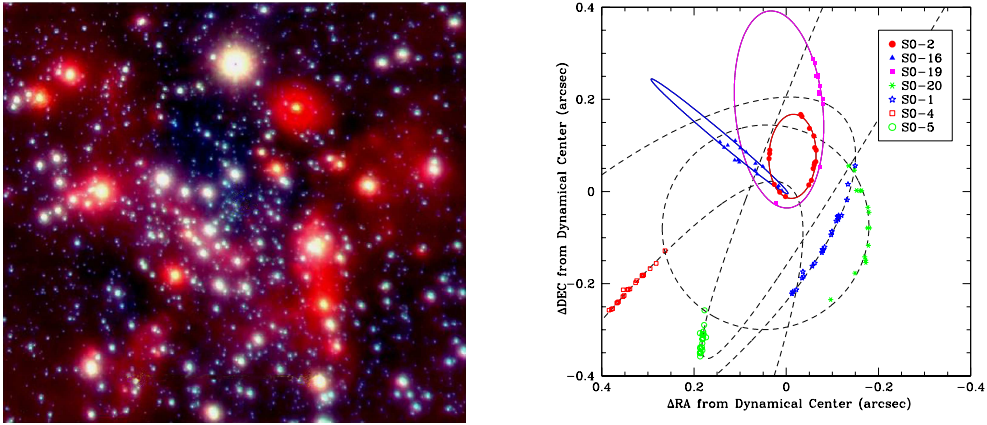}
\centering
\caption{\emph{Left} panel: a diffraction limited image of Sagittarius A$^*$ ($\sim$ 0.05 arcsec resolution)
from the 8m ESO Very Large Telescope, taken with the NACO AO-camera and an infrared wavefront sensor
at 1.6/2.2/3.7 $\mu$ m. The central black hole is located in the centre of the
box. NACO is a collaboration between ONERA (Paris), Observatoire de Paris, Observatoire
Grenoble, MPE (Garching), and MPIA (Heidelberg). \emph{Right} panel: positions on the sky as a function of time for the central stars orbiting the compact
radio source Sagittarius A$^*$. The data are from the UCLA group working with the Keck telescope.
Figures taken from \citet[][]{Genzel07} and references therein.} \label{fig|Genzel}
\end{figure}

The detection of any massive dark object at the center
of a currently inactive galaxy is performed by analyzing
the motion of gravitating bodies embedded in the
strong BH gravitational field. In any given galaxy, a BH detection can be roughly labeled in terms of $r_G$, the radius of the sphere of influence
within which the BH dominates the gravitational potential of its host galaxy. However, it is easily seen that for a $10^8$\msun, at a distance
of 20 Mpc, $r_G$ is only 0.11 arcsec \citep[e.g.,][]{Barth04}, which proves that
reliable direct dynamical BH mass measurements can only be safely
obtained in nearby galaxies. In particular, the presence of BHs in inactive
galaxies is tested through the velocity dispersion of stars, water masers, and gas
(see \citealt{FerrareseFord}). The possibly most exciting
example is the detection of a BH at the center of our own Galaxy, which
also represents one of the tightest constraints on the existence of a BH.
Monitoring has been carried out for more than 10 years on 40 stars
orbiting around the center of the Galaxy. Measurements in the infrared with adaptive
optics imaging of the 10m class telescopes have then allowed determining the three dimensional, Keplerian orbits of stars
within ten light hours of the compact radio source at the center of the Milky Way (see Figure~\ref{fig|Genzel}). A few
of them have already passed the pericenter allowing \citet[][and references therein]{Genzel07}
and \citet{Ghez03} to measure a central mass of $\sim 3\times 10^6$\msun.

BHs are currently measured in active galaxies too. The first attempts
used direct spectral fitting of the Big Blue Bump,
the optical/UV/soft X-ray continuum thermal emission from an optically thick, geometrically
think accretion disk. More secure mass determinations are
currently based on the Reverberation Mapping
technique and extrapolations of the latter (see \citealt{FerrareseFord}
for a full description of such methods).

Given the overwhelming evidence
for a strong connection between BHs and their hosts, in particular
all massive galaxies have a central massive BH, most theoretical models of galaxy formation now include
recipes to grow BHs within galaxies
\citep[e.g.,][]{Granato04,DiMatteo05,Vittorini05, Bower06,Croton06,Hopkins06,Lapi06,Malbon07,Monaco07,Menci04,Menci06,Menci08,Cook09}. However,
as explained below, models differ significantly
in the physical recipes adopted to evolve BHs, and although
sometimes the observables reproduced are the same, the
underlying physics might be widely different.

It is therefore of extreme importance to probe BH evolution
in ways independent of specific models. In this Review I will show
how along the last years, several groups have been
able to develop and refine semi-empirical techniques
to constrain BH evolution from first principles.
The Review will focus
on the latest statistical results on the BH/AGN population
and show how these information can be
interconnected to derive interesting
and firm constraints on how BHs must have evolved
along cosmic time.

This Review will therefore not discuss details of BH/AGN emission
processes (see the seminal paper by \citealt{Rees84}), nor the problem
of seeding galaxies with BHs (e.g., \citealt{HaimanQuataert}), nor the
problem of AGN unification models
\citep[e.g.,][and references therein]{UrryPadovani,Maraschi02,Elitzur08}.
On the other hand, it will cover aspects of BH evolution
which have not been properly discussed before
in a coherent framework.

In \S~\ref{sec|correlations} all the main correlations
which have been measured between BH mass and host galaxy properties will be discussed, putting more emphasis on the most recent empirical results
still not covered by previous reviews on the subject. In \S~\ref{sec|BHMF}
it will be shown how to measure the comoving number of BHs, and \S\S~\ref{sec|evolution}, \ref{subsec|BHMFevol}, and \ref{subsec|merging}
will discuss why the latter information represents an invaluable tool to understand
how BHs have grown to the population we observe today. \S~\ref{sec|clustering} will instead
discuss how quasar clustering provides many, useful and independent constrains on BH evolution. A brief overview of how current models of galaxy formation
are currently treating BH evolution, will be given in \S~\ref{sec|models}. This aspect
has not yet been systematically addressed by previous reviews
on BHs, and it is useful to show how current open problems
in galaxy-BH co-evolution models can be naturally and simply addressed
via the techniques outlined in this Review. The
main conclusions of this whole discussion will be listed in \S~\ref{sec|conclu}.

\section{The tight relation between Black Holes and their host galaxies
in the local Universe}
\label{sec|correlations}

The presence of nuclear BHs in the majority
of galaxies (or at least {\it bulges}) has become an accepted
paradigm. The nuclear kinematics of all nearby galaxies for
which data of the necessary depth and spatial resolution have been
collected, show the clear signature of central mass concentrations
beyond what can be attributed to known stellar populations. The
implied BH masses \mbh\ are found to correlate with several
global properties of the host galaxies, including the mass, $M_{\rm
bulge}$ (or luminosity $L_{\rm bulge}$), velocity dispersion, $\sigma$,
and light concentration $c_{\rm bulge}$ of the spheroidal component
\citep[e.g.,][]{Ferrarese00,Gebhardt00,Graham01,Marconi03,HaringRix},
and the mass of the surrounding dark matter halo \citep[e.g.,][]{Ferrarese02}.
In view of the clear importance of these correlations, it is worth discussing them
below in some detail before progressing further in the Review.

\subsection{The Correlation between Black Hole mass and the luminosity/mass of the bulge}
\label{subsec|MbhMstar}

\citet{Kormendy95} reviewed the status of BH demography up to that time.
Using a sample of only 8 well measured BH mass measurements they pointed out a tight
correlation, with the BH mass being 2-3\% of the stellar mass of the host.
\citet{Magorrian98} constructed 2D dynamical models for a sample of 36 nearby galaxies finding that the models required at the 95\% confidence level the mass of the BH to be linearly
correlated with the mass of their hosts as \mbh$\sim 0.006$\mstar.
Updating the catalog of BH detections, \citet{KormendyRichstone00}
concluded that the BH mass correlates much better with \emph{bulge} luminosity than with the total luminosity of the galaxy, and that BH mass correlates with the luminosity of the high-density central component in disk galaxies independent of whether that component is a classical bulge (essentially a mini-elliptical) or a ``pseudobulge'' (believed to form via inward transport of disk material).
\citet{McLureDunlop01BH} were the
first to point out that if the sample is restricted to elliptical galaxies, the
scatter is further significantly reduced. Despite the strong improvements
in bulge-bar-disk decompositions \citep[e.g.,][]{Peng02,Gadotti09,Laurikainen09},
due to the still small sample of local galaxies with secure BH mass measurements,
it is still
not clear whether the reduced scatter in ellipticals is mainly due to uncertainties
associated in determining
bulge masses in later-type galaxies, or to an intrinsic physical effect
(see also discussions in \citealt{FerrareseFord}).
\citet{McLureDunlop01BH} also showed that both active and inactive
galaxies follow similar correlations both in slope and normalization, although the intrinsic
scatter around the correlations for inactive galaxies is much lower.

\citet{Marconi03} used $K$-band images from the 2MASS database, and performed an accurate bulge/disk decomposition
for the nearby quiescent spiral galaxies with dynamically measured
masses. They confirmed a decrease of the intrinsic scatter
around the correlations when only early-type galaxies were considered.
However, their total $K$-band magnitudes were derived by extrapolating
the image profiles to very large distances on implying an average
increase of brightness of about 0.5 magnitudes \citep[e.g.,][]{Batcheldor07}.

\citet{Graham07} critically revisited all the differing \mbh-$L$ relations discussed above by McLure \& Dunlop, Marconi \& Hunt and others. He considered several previously
not properly addressed issues finding that, when these adjustments are consistently applied to all the previous samples, results in relations which predict similar BH masses. His final result in the $K$-band reads
\begin{equation}
\log (M_{\rm bh}/M_{\odot}) = (-0.37\pm 0.04)(M_K + 24) + 8.29\pm 0.08\, ,
\label{eq|GrahamMbhL}
\end{equation}
with an intrinsic scatter of 0.33 dex.

\citet{Gultekin09}
recently recomputed the relations between BH and host-galaxy bulge velocity dispersion and luminosity, using the updated sample of 49 \mbh\ measurements and also adopting 19 upper limits. Restricting to early-type galaxies, they find
\begin{equation}
\log (M_{\rm bh}/M_{\odot}) = (8.95\pm 0.11)+ (1.11\pm 0.18)\times \log (L_V / 10^{11} L_{\odot,V})\, ,
\label{eq|Gultekin}
\end{equation}
with an intrinsic scatter of $0.38 \pm 0.09$ dex.
The Authors claim their results to be insensitive to a wide range of assumptions about the measurement uncertainties and the distribution of intrinsic scatter. They also discuss
how restricting the sample to the objects with resolution of the BH's sphere of influence might bias the results, a conclusion at variance with previous
studies \citep[e.g.,][]{Marconi03,FerrareseFord}. 


\citet{HaringRix} approached the correlation between BH mass
and host stellar mass in a way independent of photometric measurements,
i.e., utilizing a total of 30 galaxies with bulge masses derived through Jeans modeling or adopted from dynamical models in the literature. They found \mbh$\sim 1.4\times 10^{-3}$\mbulge\ with an observed scatter of $\lesssim 0.30$ dex, a fraction of which attributed to measurement errors. Their results confirm and refine the previous work of \citet{Marconi03}.

In sum, all of the most recent and detailed investigations
have settled to define a closely linear correlation between BH mass and host galaxy mass/luminosity over several decades, with a factor of proportionality of $2\times 10^{-3}$, e.g., \mbh$\simeq 2\times 10^{-3}$\mbulge.

\subsection{The black hole mass velocity dispersion correlation}
\label{subsec|Mbhsigma}

The tight correlation between BH mass and the velocity dispersion \sis\ of the host
galaxy is probably the correlation which received most
attention in the literature as it provides a direct proof of the
strong link between BH mass and potential well of the galaxy.
It was first pointed out by \citet[][]{Ferrarese00,Gebhardt00,Tremaine02}, although with
some discrepancies in the basic results (see also \citealt{Marconi04}), and we
refer to those papers and to the review by \citet{FerrareseFord} for extensive
discussions on this topic.

One of the latest determinations of the \mbh-\sis\ obtained by \citet{Tundo07} yields the result
 \begin{equation}
 \log (M_{\rm bh}/M_{\odot})=(8.21\pm 0.06)+(3.83\pm 0.21)\log \left(\sigma/200{\rm \, km\, s^{-1}} \right) \, ,
 \label{eq|Tundo}
 \end{equation}
with an intrinsic scatter of $0.22\pm 0.06$ dex.
These Authors also point out that the \sis-$L$ relation
of local early-type galaxies, including those having BH mass measurements,
is different from the average \sis-$L$ relation of early type galaxies
calibrated in the Sloan Digital Sky Survey (SDSS). This difference, they conclude,
might bias the local BHMF. \citet{Lauer07demo} also
stress that the curvature in the \sis-$L$
relation for massive early-type galaxies implies that
the \mbh-\sis\ relation is unlikely to be a
single power law. On similar lines, \citet{Wyithe06}
suggested through a Bayesian analysis that a quadratic
fit is a better description of the correlation between BH mass and \sis\ in the Tremaine
et al. (2002) sample. \citet{Hu08} reinvestigated in detail the \mbh-\sis\ relation, finding
similar results to Eq.~(\ref{eq|Tundo}) for the ellipticals, but also
pointing out that pseudo-bulges and core early-type galaxies seem to
show significant departures from it,
a possible signature of their different formation histories.

\subsection{Redshift Evolution of the correlations}
\label{subsec|zEvolutionCorrelations}

Both the local characterization, and the cosmic evolution of
the above mentioned scaling relations are subject to intense observational studies \citep[e.g.,][]{Shields06,Peng06,Lauer07bias,Treu07,ShankarMsigma}, as they
provide powerful constraints on theoretical models for the
co-evolution of galaxies and BHs \citep[e.g.,][]{Granato04,Vittorini05,Hopkins06,Croton06,DeLucia06,Cav07,Monaco07,Marulli08}.

\citet{Treu07} and \citet{Woo08} have randomly compiled
from the SDSS Data Release 4 a sample of about 20 Seyfert
galaxies in the redshift range $0.37 \lesssim z \lesssim 0.57$.
\citet{Shen08bias} estimated
the \mbh-\sis\ relation for a larger sample of active galaxies up to
$z = 0.452$. While the latter claim that no significant evolution
in the \mbh-\sis\ relation is detectable from their sample, \citet{Woo08}
confirm the results by \citet{Treu07} that a
significant increase of $\sim 0.2$ dex in BH mass at fixed velocity
dispersion must occur within $z = 0$ and $z \sim 0.5$. However, systematic uncertainties may
affect these estimates; for example, as also discussed by \citet{Woo08},
especially in galaxies with lower BH mass the
host galaxy contribution to the 5100 {\AA} luminosity may lead
to an overestimation of the true BH mass.

\citet{Peng06} collected a sample of 31 lensed and 18
unlensed AGNs at redshifts $z> 1.7$.
They measured rest-frame R-band luminosities from H-band
fluxes and BH masses by applying virial relations based on
emission line widths. They found that the BH-to-host galaxy
luminosity at $z \sim 2$ is about the same as that at $z \sim 0$.
Therefore,
once the observed rest-frame luminosity has dimmed through
substantial evolution to $z \sim 0$, at fixed BH mass the ratio
BH-to-host
luminosity grows significantly, and the resulting BH-luminosity
normalization is several times higher than the local one.
Similar results were derived by \citet{McLure06}, who measured the
BH-to-host galaxy mass ratio in a sample of radio-loud AGNs
in the redshift range $0 <z< 2$ finding \mbh$/$\mstar$\, \propto (1 + z)^2$.
\citet{Shields06} found that the CO emission lines in a
sample of $z> 3$ quasars are very narrow, suggesting bulge
masses about an order of magnitude lower than measured in the
local Universe, at fixed BH mass (see also \citealt{Coppin08}).

\citet{Lauer07bias} have discussed several
possible biases which may seriously affect these findings. At
high redshifts a sample will be biased toward the most luminous
AGNs and more massive BHs. Given the observed scatter in the
local relations, especially significant in the BH–host luminosity
relation, these massive BHs will be preferentially associated
with the less massive but more numerous galaxies yielding
a false signal of evolution.

Several approaches have been
pursued in the literature to address
the issue of redshift evolution in the relations
adopting methodologies which minimize the impact of observational biases.
\citet{MerloniSFR} compared the accreted BH mass
density in AGNs with the cosmological global star formation
rate density (see also \citealt{Haiman04}). Although their
conclusions depend on additional assumptions about the fraction
of the star forming galaxies which are linked to BH growth at a
given redshift, they were able to rule out a strong
evolution at a high confidence level; we will further discuss their
results below. \citet{Hop06limit} also found, through a model-independent integral constraint
based on comparing the AGN density with
the luminosity and mass functions in different bands within the
redshift range $z = 0 \div 2$, very significant evidence for
an extremely mild evolution.
\citet{ShankarMsigma} utilized the local velocity dispersion function (VDF) of spheroids, together with their inferred age distributions,
to predict the VDF at higher redshifts ($0 <z < 6$).
They then compared via
a redshift-dependent \mbh-\sis\ relation
the cumulative
BH mass density inferred from the AGN luminosity function
with that associated with the VDF at each redshift.
This comparison, insensitive to the assumed duty cycle or Eddington ratio of quasar activity, yielded a very mild redshift evolution, in agreement with previous studies.

\subsection{Other correlations between Black Holes and host galaxies}
\label{subsec|OtherCorrelations}

With a more numerous sample of galaxies with updated estimates of galaxy distances, BH masses, and S\'{e}rsic indices $n$, \citet{Graham07BHMF} found evidence for
a quadratic \mbh-$n$ relation, of the type
 \begin{equation}
 \log (M_{\rm bh}/M_{\odot})=(7.98\pm 0.09)+(3.70\pm 0.46)\log(n/3)-(3.10\pm 0.84)[\log(n/3)]^2\, ,
 \label{eq|GrahamMbhSersicRelation}
 \end{equation}
 with an intrinsic scatter of
 $0.18^{+0.07}_{-0.06}$ dex and a total absolute scatter of 0.31 dex. When the quadratic relation is extrapolated, it predicts BHs with masses of $\sim 10^3\, $\msun\ in $n=0.5$ dwarf elliptical galaxies, and an upper bound on the largest BH masses in the local Universe equal to $\sim 1.2\times 10^9$\msun. This relation
 when adopted to compute the statistics of local BHs, yields quite different results with respect
 to other methods, as discussed in \S~\ref{sec|BHMF}.

\citet{Marconi03} explored the addition to the \mbh-\sis\
relation of the parameter \re\ (the effective radius)
to create a ``fundamental plane'' for BHs.
From their 27 ``Group 1'' galaxies, they constructed a relation
between \mbh\ and and \re\sis$^2$ (proportional to the virial
bulge mass), which resulted in an intrinsic dispersion
of $\sim 0.25$ dex, comparable to the one in the \mbh-\sis\ relation.
Allowing the exponents on the \re\ and \sis\ terms
to vary independently, \citet{Hop07FP} used the
same 27 Group 1 galaxies from \citet{Marconi03} along with some
updated measurements, to find the best-fit relation
\begin{eqnarray}
\log (M_{\rm bh}/M_{\odot}) = (8.33 \pm 0.06) + (0.43 \pm 0.19) \log[r_e/{\rm 3\, kpc}] \nonumber\\
+ (3.00 \pm 0.30) \log[\sigma/200 {\rm \, km\, s^{-1}}]\, ,
\label{eq|HopkinsFP}
\end{eqnarray}
with an intrinsic scatter of 0.21
dex and a total scatter of 0.30 dex (see \citealt{Graham08FP}).

\citet{Barway07} made the
interesting claim that a combination of two photometric
parameters, namely the effective radius \re\ and the mean
effective surface brightness $\mu_e = -2.5 \log I_e$, can be
used to predict BH masses with a higher degree of
accuracy than single quantities such as luminosity or velocity dispersion.
The optimal (B-band) solution using all 18 data points
as reported by \citet{Graham08FP} is
\begin{eqnarray}
\log (M_{\rm bh}/M_{\odot})=(8.18 \pm 0.09) + (3.15 \pm 0.33) \log [r_e/{\rm 3\, kpc}]\nonumber \\
-(0.90 ± 0.18)[\mu_{e,B} - 21.0]\, ,
\label{eq|BarwayKembhavi}
\end{eqnarray}
with a total scatter of 0.32 dex.


The hypothesis of a BHFP which might be more fundamental
than the \mbh-\sis\ relation is obviously of great interest. However,
later studies have questioned the actual existence of a BHFP.
Graham (2008) showed that the barred early-type galaxies contained in the sample
systematically deviate from the \mbh-\sis\ relation, causing the
residuals about the \mbh-\sis\ relation to correlate with \re\ giving rise to the apparent BHFP
(see, however, discussion in \citealt{Younger08BH}).
\citet{AllerRichstone07} investigate a variety of correlations between
 bulge properties and BH mass, concluding that the strongest
 correlation exists between the bulge gravitational binding energy $E_g$,
as traced by the stellar light profile, and the BH mass, of the type
\mbh$\propto E_g^{0.6}$. They, however, do not find evidence for a significant
correlation between between BH mass, \sis, and \re\ in early-type galaxies, with their best-fit
relation being \mbh$\propto \sigma^{3.73}$\re$^{0.05}$.

\citet{Kormendy09} have adopted accurate measurements
of the stellar ``light deficit'' $L_{\rm def}$ associated with the cores
of a sample of elliptical galaxies characterized by a surface
brightness covering a quite large dynamic range. They find, in agreement with several
previous studies (e.g., \citealt{Merritt06}, and references therein), that $L_{\rm def}$
correlates well with the BH mass, with a similar, or even smaller
scatter than the one in the \mbh-\sis\ relation. They claim
that this might be associated to dry mergers events, during which energy can be efficiently
transferred to the central stars via the anisotropic gravitational radiation emitted after the coalesce of the binary BHs \citep[e.g.,][and references therein]{Sesana09}.
Finally, \citet{Seigar08} claimed another
correlation between the spiral arm pitch angle (a measure of the tightness of spiral structure) and the BH mass in the nuclei of disk galaxies.

\section{The local mass function of Black Holes}
\label{sec|BHMF}

It is now widely accepted that knowing the exact number density
of galaxies in all bins of stellar
mass (or other property) is fundamental
to discern among successful models of galaxy formation.
For example (see \S~\ref{sec|models}), most semianalytic models (SAMs) of
galaxy formation now agree that some sort of strong
stellar and/or AGN feedback must have been responsible
for shaping the exponential fall-off of the stellar mass function
at high stellar masses.

For similar reasons, it is essential to define the BH mass function (BHMF),
i.e., the comoving
number density of BHs per bin of mass in the local Universe.
Any model of galaxy formation and evolution
which attempts to provide a complete picture of galaxy
evolution, must also address the problem of birth and growth
of the massive BHs in galaxies, explain the nature
of their tight relationship with their hosts, and
predict their correct number and mass at $z=0$.

The standard procedure (first pioneered by \citealt{Salucci99} and \citealt{Ferrarese02Proc})
in calculating the BHMF assumes that all
galaxies with a significant bulge component host a BH, the mass
of which can be
inferred from the BH$-$galaxy scaling relations
discussed in \S~\ref{sec|correlations}.
Knowledge of a local galaxy luminosity,
velocity dispersion or light concentration function, combined with
the corresponding BH scaling relation, will then lead to a BHMF
through a simple change of variables. The first derivations of the BHMF
\citep[e.g.,][]{Salucci99,Ferrarese02Proc,AllerRichstone02,MD04,Marconi04,Shankar04,Tundo07,Graham07BHMF}
combined the galaxy velocity
dispersion function (VDF) with the \mbh-\sis\
relation, and the spheroid luminosity function with the
\mbh-\mbulge\ relation.

Here it is provided
a brief compendium (see \citealt{SWM}) of the BHMFs
derived from applying different BH mass estimators. The aim here
is to provide the reader
with a compact overview of the
most important steps required to generate a BHMF.

The BHMF is computed by converting
the galaxy distribution $\Phi(y)$, which expresses the measured galaxy
number density as a function of a given variable $y$ (e.g., the stellar velocity dispersion
or bulge luminosity), into a BHMF by adopting the corresponding
empirically measured \mbh-$y$ relation. The conversion is then
performed via the convolution
\begin{equation}
\Phi(M_{\rm bh})=\int \Phi(y) \frac{1}{\sqrt{2\pi
\eta^2}}\exp\left[-\frac{\left(M_{\rm bh}-\left[a+by \right]
\right)^2}{2\eta^2}\right]dy\nonumber
    \label{eq|BHMFscatter}
\end{equation}
which takes into account the intrinsic scatter $\eta$ in
the \mbh-$y$ relation, derived from the direct data fitting
\citep[e.g.,][]{Tremaine02,Graham07}.

The BHMF can be derived by coupling the statistical information on local luminosity functions
of galaxies with relationships among luminosity
(or related quantities such as stellar mass and velocity dispersion)
and the central BH mass (see \citealt{Salucci99}).
Because the BH mass correlates with the luminosity and velocity
dispersion of the bulge stellar population, it is necessary to first separate the luminosity functions
for different morphological types, that have different bulge-to-total
luminosity ratios. It is convenient to use galaxy luminosity functions derived in
bands as red as possible, where the old bulge stellar populations are
more prominent.

The spheroidal luminosity function $\Phi(L_{\rm bulge})$ for
lenticular galaxies can be computed from the luminosity function of
early-type galaxies, $\Phi_e(L)$, as
\begin{equation}
\Phi(L_{\rm bulge})=\frac{f_{\rm S0}}{f_{\rm E}+f_{\rm
S0}}\Phi_e(b\times L)\, ,
    \label{eq|fsph}
\end{equation}
(\citealt{YT02}), where $f_{E}$ and $f_{S0}$ (which can be taken, for example,
from Table 1 of \citealt{Fukugita98}) are the numerical fractions of elliptical and
lenticular galaxies, respectively, to the total galaxy population,
and $b=L_{\rm bulge}/L$, is the luminous bulge-to-total ratio for $S0$ galaxies. The
contribution of ellipticals to the spheroidal LF can be computed by
simply setting $b=1$ and replacing $f_{\rm S0}$ with $f_{\rm E}$ in
equation~(\ref{eq|fsph}). Similarly, one can compute the luminosity
function of bulges of spirals of different Hubble type by replacing
$\Phi_e(L)$ in equation~(\ref{eq|fsph}) with the late-type galaxy
luminosity function, and plugging in the appropriate $f$-weights for
later type galaxies.

A somewhat different approach was recently adopted by \citet{Graham07BHMF}.
First a BH mass via Eq.~(\ref{eq|GrahamMbhSersicRelation}) and the appropriate
statistical weight were associated to each spheroid in the Millennium Galaxy Catalog (\citealt{Driver07}).
The latter was computed as the space density of
of the appropriate spheroid type in the specified color and luminosity
intervals, divided by the number of galaxies which contributed to that interval.
The BHMF was then computed separately for early and late-type galaxies
by summing the
distribution of appropriate BH masses times their weights
(see \citealt{Graham07BHMF} for further details).

When computing the BHMF from the velocity dispersion function
of galaxies, instead, no correction for bulges
is usually applied, at least for the bulge-dominated
galaxies where the velocity dispersion \sis\ should
be a good indicator of the true \sis\ of the bulge
component (see discussion in \citealt{FerrareseFord}, and references
therein). Several groups therefore \citep[e.g.,][]{Marconi04,Shankar04},
adopted the \citet{Sheth03} early-type velocity dispersion function,
directly converted into a BHMF via the \mbh-\sis\ relation.
\citet{Shankar04} also adopted
the velocity dispersion function computed via the bivariate equation
\begin{equation}
\Phi(\sigma_i)=\sum_{j}p_{ij}\Phi(L_j)\, ,
    \label{eq|bivar}
\end{equation}
where $p_{ij}$ is the fraction of sources in the \citet{Bernardi03} and \citet{Sheth03} sample with velocity dispersion
$\log \sigma_i$ and luminosity $L_j$, normalized to the total number
of sources with luminosity $L_j$, and $\Phi(L_j)$ is the input
luminosity function.
The contribution of late-type galaxies to the velocity
dispersion function is more difficult
to assess. \citet{Sheth03} converted the late-type galaxy
luminosity function into a velocity dispersion function by adopting a mean
luminosity-velocity dispersion relation plus a model to correct for
galaxy orientation. The latter result was then adopted
by \citet{SWM} to compute the contribution
of the bulges of spirals to the BHMF.

\begin{figure}
\includegraphics[angle=00,scale=0.8]{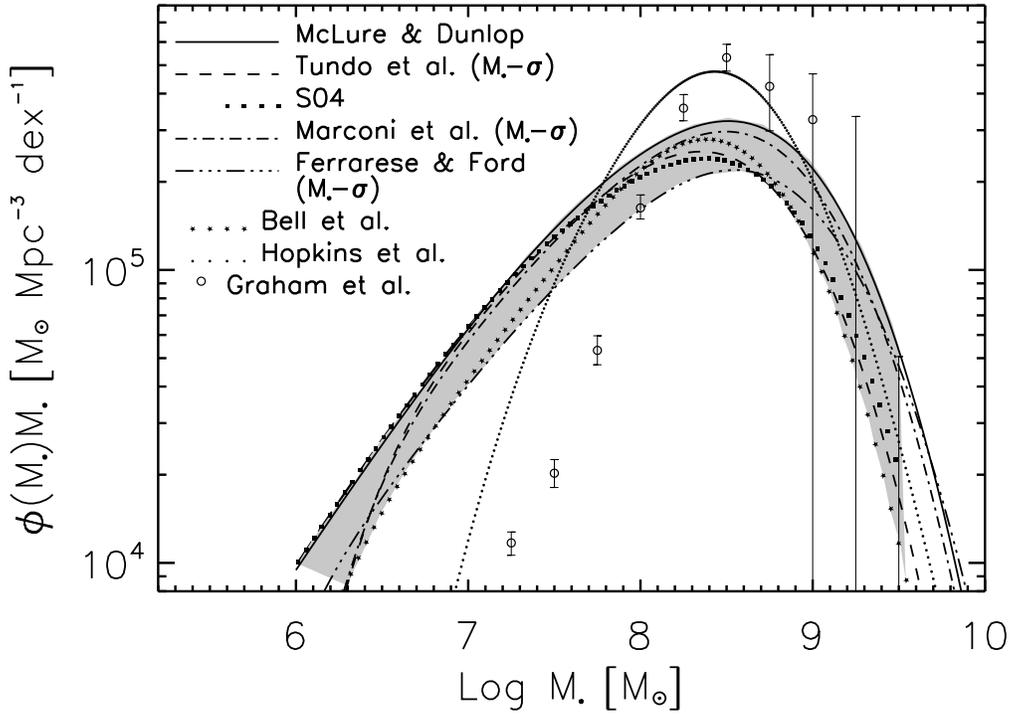}
\caption{Comparison among estimates of the local black hole
mass function taken from \citet{SWM}. \emph{Lines} show estimates using various calibrations of the $M_{\rm bh}-L_{\rm sph}$,
$M_{\rm bh}-\sigma$, or $M_{\rm bh}-M_{\rm bulge}$ relations,
assuming a 0.3-dex intrinsic scatter in all cases. The \emph{grey band}
encompasses the range of these estimates. The \emph{thick dotted} line shows the
determination of the early-type black hole mass function
by \citet{Hop07FP} using the black hole
``fundamental plane'', and \emph{open circles} show the determination of \citet{Graham07BHMF}
using the relation between black hole mass and S\'{e}rsic index.} \label{fig|LocalMFs}
\end{figure}

Figure~\ref{fig|LocalMFs} compares the local BHMFs obtained
from using different relations between BH mass and host galaxy
properties. Each line in the Figure corresponds
to a different BH mass estimator, as listed in the legend
(we defer the reader to \citealt{SWM} for further details).

The grey band of Figure~\ref{fig|LocalMFs}
can be considered as representative of the mean and the systematic uncertainties of present estimates of the BHMF.
The integrated mass density of the local BH population is $\rho_{\rm bh}=(3.2-5.5)\times 10^5\, {\rm M_{\odot}\, Mpc^{-3}}$ (for $h=0.7$).
Figure~\ref{fig|LocalMFs} also presents two additional estimates of the BHMF.
The dotted curve shows the estimate, for early-type galaxies only,
by \citet{Hop07FP}, based on the BHFP (Eq.~[\ref{eq|HopkinsFP}]), which is
in reasonable agreement with other determinations of the BHMF at the high mass end.
The disagreement at the low mass end is due to the fact that
the contribution from the bulges of spirals is missing from their
estimate. Open circles
show instead the estimate by \citet{Graham07BHMF} based
on the correlation between BH mass and the S\'{e}rsic index $n$ (Eq.~[\ref{eq|GrahamMbhSersicRelation}]).
The latter includes a complete census of all galaxy types, and therefore
the strong disagreement at the low mass end is a manifestation
of a real mismatch in the results. This difference might be
caused by both the use of a different relation (Eq.~[\ref{eq|GrahamMbhSersicRelation}]),
and, as explained above, a different method with respect to that outlined in Eq.~(\ref{eq|fsph}), although
this is difficult to assess a posteriori.
Nevertheless, it is interesting to note that all estimates agree in providing
a similar peak in the BH mass distributions, which is
important to constrain BH accretion models, as we will discuss below.

\subsection{Black Holes in Later type galaxies}
\label{subsec|BHsInLaterTypeGalx}

One important
caveat in determining the overall census
of local BHs, is the actual
presence of BHs in bulgeless galaxies.
In the previous section, the assumption has been made that all
galaxies with a bulge component possess a central BH.
However, there is no firm observational evidence to support this
claim. On one side, bulgeless AGNs are not uncommon \citep[e.g.,][]{Gebhardt01,Satyapal08},
suggesting that current BHMF
estimates might be missing the contribution of a BH population in
late type spiral galaxies (although, based on current estimates, these
BHs are $1-3$ orders of magnitude less massive than those
discussed in this Review). On the other, the
ubiquitous presence of BHs, especially in low luminosity dwarf
ellipticals, has not been proven: for instance, it is questionable
whether a BH exists in the E5 galaxy NGC 205 (\citealt{Valluri05}). A
number of recent studies, mostly with the Hubble Space Telescope,
have shown that the majority of galaxies, regardless of Hubble type,
host morphologically distinct stellar nuclei \citep[e.g.,][]{Cote06,Ferrarese06,Boeker07}, the only exception (as a
class) being bright galaxies with $M_B\lesssim -20.5$,
although cases of galaxies with both stellar nuclei and BHs do exist \citep[e.g.,][]{Seth08}.

Evidence for BHs in smaller galaxies also
come from the Chandra X-ray observations of \citet{Gallo08},
who showed that a large fraction
of early-type galaxies
in the Virgo Cluster (AMUSE-Virgo)
have a significant X-ray emission, consistent
with what expected from accretion onto a central BH.
\citet{Satyapal08} reported the discovery of AGN
activity in several local spiral galaxies, and
concluded from their estimation of bolometric
luminosity, that the average mass for the BHs
residing in these galaxies should be around $4\times 10^3$\msun, if they emit
at the Eddington limit.
\citet{Ghosh08} found secure X-ray AGN
detections in several nearby, face-on spiral galaxies.

It remains still open the major challenge of understanding how BHs have
grown and whether they followed the assembly history of their
host galaxies (see \S~\ref{sec|models}).
It is discussed in the following sections how, in ways independent
of specific models, important
constraints on how BHs have actually evolved within their
host galaxies and dark matter halos can be derived by using
global information on the statistics of local BHs, the statistics
of active galaxies and their correlated clustering strengths
at different luminosities and redshifts.

\section{The integrated mass density}
\label{sec|evolution}

%
%


The classic \citet{Soltan} argument relates the integrated
BH density to the integrated emissivity of the AGN
population, setting interesting constraints on the average radiative
efficiency $\epsilon$ of BHs and its possible dependence on mass and/or redshift.
If the average efficiency of converting accreted mass
into bolometric luminosity is $\epsilon \equiv L/\dot{M}_{\rm
inflow}c^2$, where $\dot{M}_{\rm inflow}$ is the mass accretion
rate, then the actual accretion onto the central BH is
$\dot{M}_{\rm bh}=(1-\epsilon)\dot{M}_{\rm inflow}$, where the factor $1-\epsilon$ accounts for the fraction of
the incoming mass that is radiated away instead of being added to
the BH (e.g., \citealt{YT02}).
The rate at which mass is added to the BH mass function
is then given by
\begin{equation}
\frac{d\rho_{\rm bh}}{dt}=\frac{1-\epsilon}{\epsilon
c^2}\int_0^{\infty}\Phi(L)Ld\log L\, ,
    \label{eq|soltan}
\end{equation}
with $\Phi(L)$ the bolometric AGN luminosity function.

\begin{figure}
\includegraphics[width=13truecm, scale=1.4]{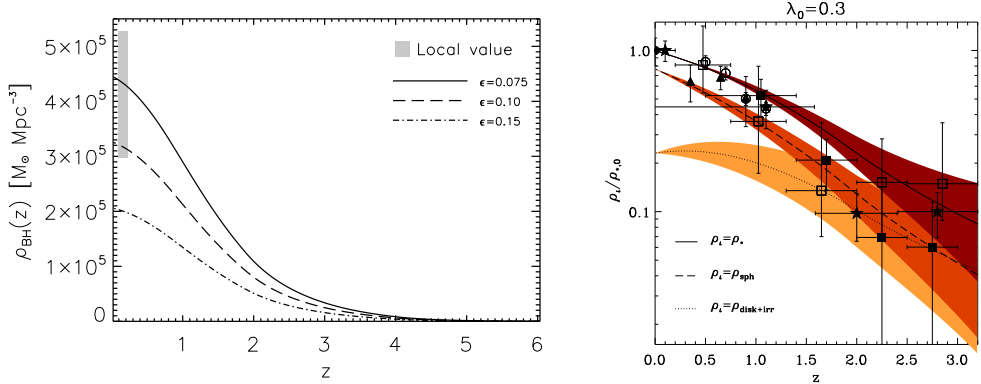}
\caption{\emph{Left panel}: integrated accreted mass on black holes as a function
of redshift following Eq.~(\ref{eq|soltan}), for different values
of the radiative efficiency $\epsilon$, as labeled. The grey stripe at $z\sim 0$ indicates
the systematic uncertainties in determining the local BH mass function (see Figure~\ref{fig|LocalMFs}). The match between the local and accreted mass densities provides interesting constraints
on the mean radiative efficiency of BHs. \emph{Right panel}: evolution in the best-fitting stellar mass
density as a function of redshift derived from the mean BH accretion history rescaled
by a redshift-dependent constant (\emph{solid} line), compared with different data sets as collected in \citet{MerloniSFR}. Figure taken from \citet{MerloniSFR}.}
\label{fig|Soltan}
\end{figure}

Here I show a simple plot which conveys the basic
idea of \citet{Soltan}'s argument, while I refer to
other papers for a more detailed discussion on the topic
\citep[e.g.,][]{Marconi04,MerloniSFR,Shankar04,Tamura06,Hop07,MerloniHeinz08,Silverman08LF,SWM}. Figure~\ref{fig|Soltan} shows the integrated accreted mass on BHs as a function
of redshift following Eq.~(\ref{eq|soltan}), for different values
of the radiative efficiency $\epsilon$, as labeled. The grey stripe at $z\sim 0$ indicates
the systematic uncertainties in determining the local BHMF (see Figure~\ref{fig|LocalMFs}). It is clear from the Figure that the match between the local and accreted mass densities constrains the mean radiative efficiency to be $\epsilon \lesssim 0.1$. However, still significant uncertainties on the bolometric correction on one side \citep[e.g.,][]{Elvis94,Marconi04,Hop07,Vasudevan07,Vasudevan09}, and systematic uncertainties on the local BHMF on the other \citep[e.g.,][and references therein]{SWM}, prevent a more precise computation of the radiative efficiency (see also discussions in \citealt{SWM} and \citealt{YuLu08}).

Extended versions of the \citet{Soltan} argument have been successfully
applied to constrain other aspects of BH evolution. An upper limit
to the degree of evolution of the \mbh-\sis\ relation has been
set by \citet{ShankarMsigma} by comparing the BH mass function associated with the
velocity dispersion function at each redshift $z_c$ (corrected for galaxies with age $z<z_c$), and the accumulated total BH mass density extracted from Eq.~(\ref{eq|soltan}); these Authors reach
similar conclusions as \citet{MerloniSFR}, \citet{Hop06limit}, and \citet{SWM} who used somewhat different methods. Some of these studies have also shown that the inferred average
history of BH growth apparently parallels the cosmological star formation rate, at least in spheroids, at nearly all cosmological times, suggesting that the two processes may well be
intimately linked and that, on average, BHs and galaxies may coevolve.
The right panel of Figure~\ref{fig|Soltan} shows the result of \citet{MerloniSFR}. These Authors compared all the available observational
data on the redshift evolution of the total stellar mass (data points in the Figure), with the integrated mass evolution of BHs, estimated from the AGN luminosity function (solid line).
Assuming that the ratio of the BH mass
density to the stellar mass in spheroids evolves as $(1 + z)^{\alpha}$, they found that the
match to the data requires a mild, although significant, evolution
in the BH-spheroid mass ratio with $\alpha\simeq 0.5\pm 0.2$; this implies slightly
larger BH masses at fixed host stellar mass at $z>0$,
in agreement with the evolution inferred by \citet{ShankarMsigma} for the \mbh-\sis\ relation.
Also shown in the same Figure are the relative contributions
from early-type (dashed line) and late-type galaxies (dotted line),
with their corresponding 1-\sis\ confidence intervals.
\citet{MerloniHeinz08}, \citet{Shankar08Cav}, and \citet{CattaneoBest} have also
used the integrated emissivity of radio sources to constrain the kinetic efficiency of BHs, in the hypothesis that radio power is correlated with the kinetic power of the jets.

\section{The evolved Black Hole mass function}
\label{subsec|BHMFevol}

Important additional insight into the overall evolution
of BHs can be gained by comparing not only the
total mass densities, but also the \emph{shapes}
of the mass functions (\citealt{Salucci99}).
In fact, the comparison between the predicted mass distribution
implied by the AGN luminosity function (Eq.~[\ref{eq|conteq}]), and the local BHMF can yield
further interesting constraints on the mean Eddington ratio \lam\ at which
BHs are accreting at.

The predicted mass function $n(M_{\rm bh},t)$ (in units of ${\rm Mpc^{-3}\, M_{\odot}^{-1}}$) implied by the bolometric AGN luminosity function
can be derived via a continuity equation argument
\citep[e.g.,][]{Cavaliere71,SmallBlandford,YT02,Marconi04,YuLu04,YuLu08,Shankar04,Hop07,SWM}
\begin{equation}
\frac{\partial n}{\partial t}(M_{\rm bh},t)=-\frac{\partial (\langle
\dot{M}_{\rm bh}\rangle n(M_{\rm bh},t))}{\partial M_{\rm bh}}\,,
    \label{eq|conteq}
\end{equation}
where
\begin{equation}
\langle \dot{M}_{\rm bh}\rangle=\int \, S(M_{\rm
bh},z,\lambda) \, \lambda \, d\lambda \, \, M_{\rm bh}/t_E\, ,
\label{eq|meanMbh}
\end{equation}
is the mean
accretion rate (averaged over the active and inactive populations)
of BHs of mass \mbh\ at time $t$, and $\lambda=L/L_{\rm Edd}$ the Eddington ratio, with
$L_{\rm Edd}=1.26\times 10^{38} (M_{\rm bh}/M_{\odot})\, $\ergs\ the Eddington luminosity
(i.e., $\lambda \propto L/M_{\rm bh}$).
Eq.~(\ref{eq|conteq}) states that the average growth rate of all
BHs is proportional to the function $S(M_{\rm bh},z,\lambda)$, i.e.,
the fraction of BHs of mass $M_{\rm bh}$ active at redshift $z$ and
accreting at the Eddington rate \lam. This evolution is equivalent
to the case in which every BH constantly grows at the mean accretion
rate $\langle \dot{M}_{\rm bh}\rangle$. In practice, individual
BHs turn on and off, and there may be a dispersion in
$\dot{M}_{\rm bh}$ values, but the mass function evolution depends
only on the mean accretion rate as a function of mass.

Eq.~(\ref{eq|conteq}) can be further generalized to any
distribution
\begin{equation}
S(M_{\rm bh},z,\lambda)=p(\lambda,z)\Delta(M_{\rm bh},z)\, ;
    \label{eq|separatevariable}
\end{equation}
this assumes all active BHs at redshift $z$ to share the same
mean Eddington ratio distribution $p(\lambda,z)$ (normalized to unity),
and $\Delta(M_{\rm bh},z)$ is the duty cycle, i.e., the \emph{total} fraction of active
BHs at redshift $z$ and mass $M_{\rm bh}$, neglecting for simplicity
any further mass-dependence of the Eddington ratio distribution $p(\lambda,z)$.
A physically consistent model must have $\Delta(M_{\rm bh},z)\le 1$ for all \mbh\ at
all times.

In models where all BHs grow with a single value of
the Eddington ratio $\bar{\lambda}$, $p(\lambda,z)$ is just
a Dirac delta-function $p(\lambda,z)=\delta(\lambda-\bar{\lambda})$, while the duty cycle
is simply the ratio of the luminosity and mass functions,
\begin{equation}
\Delta(M_{\rm bh},z)=\frac{\Phi(L,z)}{\Phi_{\rm bh}(M_{\rm bh},z)}\big|_{L\propto \bar{\lambda}M_{\rm bh}}\, ,
    \label{eq|P0general}
\end{equation}
where $\Phi_{\rm bh}(M_{\rm bh},z)=n(M_{\rm bh},z)M_{\rm bh}\ln(10)$, and $\Phi(L,z)$ is in units of ${\rm Mpc^{-3}}$ $\, {\rm dex^{-1}}$.

Once an initial condition is specified, the solution to Eq.~(\ref{eq|conteq}) can be performed
either iteratively, updating the duty cycle computed from the AGN luminosity
function at the appropriate timestep (Eq.~[\ref{eq|P0general}]) or
in simple integral forms, at least in the case of single-\lam\ models (see, e.g., \citealt{YuLu04}).
On adopting a broad input $p(\lambda,z)$ in which different fractions
of BHs of the same mass are allowed to accrete at different
values of the Eddington ratio \lam, Eq.~(\ref{eq|conteq}) can be efficiently solved iteratively.
At any given time, given the observed AGN luminosity function $\Phi(L,z)$, the underlying
BH mass function $\Phi_{\rm bh}(M_{\rm bh},z)$,
and Eddington ratio distribution $p(\lambda,z)$, the duty cycle $\Delta(M_{\rm bh},z)$ is found fully specified by the
equality \citep[e.g.,][]{SteedWeinberg}
\begin{equation}
\Phi(L,z)=\int \, p(\lambda,z) \, \Delta(M_{\rm bh},z) \, \Phi(M_{\rm bh},z)\, d\log M_{\rm bh}\, ,
\label{eq|PhiLLambda}
\end{equation}
given the input AGN luminosity function $\Phi(L,z)$ and the Eddington ratio distribution $p(\lambda,z)$.

\begin{figure}
\includegraphics[width=13truecm,scale=1.0]{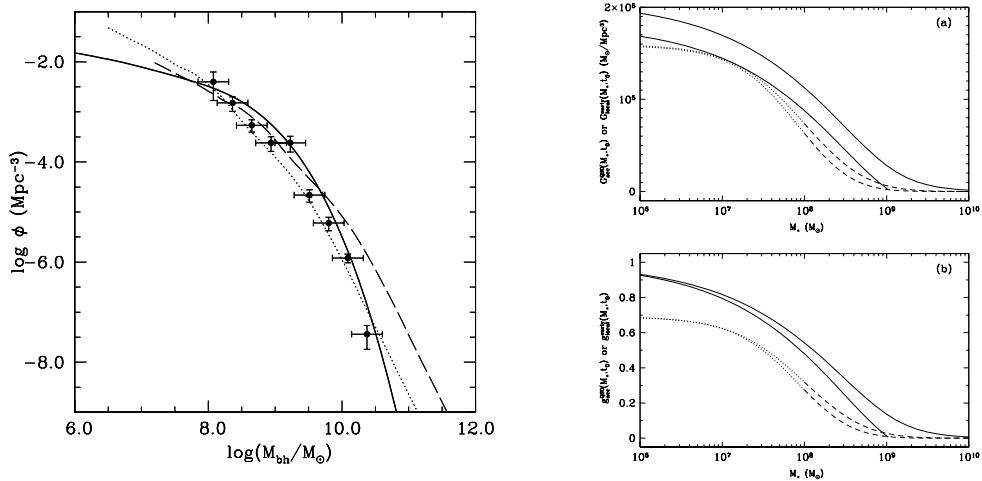}
\caption{\emph{Left panel}: predicted mass function from Eq.~(\ref{eq|conteq}) as derived from \citet{Salucci99} assuming $\lambda = 1$ (\emph{dotted} line) and $\lambda = 0.2$ (\emph{dashed} line). The \emph{solid} line is their estimate of the local black hole
mass function derived from the \mbh-\mstar\ relation, while the filled symbols refer to their estimate of the local mass function derived from radio data. Figure taken from \citet{Salucci99}. \emph{Right panel}: cumulative black hole mass function (\emph{thick solid} line) computed
from Eq.~(\ref{eq|conteq}) by \citet{YT02} assuming $\lambda=1$ and $\epsilon/(1-\epsilon)=0.1$ compared to local estimates
of the local black hole mass function shown with dotted and dashed lines (see \citealt{YT02} for details). The \emph{thin solid} line is their estimate of the predicted
black hole mass function cutting out quasars radiating above a luminosity $L=L_{\rm Edd}(10^9\, {\rm M_{\odot}})$.
The lower panel shows the same quantities as in the upper panel normalized to unity.}
\label{fig|BHMFsLiterature}
\end{figure}

Figure~\ref{fig|BHMFsLiterature} shows the results from the first attempts to match the local BHMF with the
BH mass function as predicted from Eq.~(\ref{eq|conteq}) with the assumption that all BHs accrete
with a single value of the Eddington ratio. The left panel shows the result worked out
by \citet{Salucci99}, with the dotted and dashed lines referring to the predicted mass
function for two values of $\lambda=1, 0.2$, respectively, compared with their
estimates of the local BHMF (solid line and filled symbols). The right panel shows instead a similar comparison
by \citet{YT02}, where the solid lines show the cumulative BH mass functions obtained assuming $\lambda=1$ and $\epsilon/(1-\epsilon)=0.1$, compared to their estimates
of the local BHMF shown with dotted and dashed lines. These preliminary works
found, in agreement with the \citet{Soltan} argument discussed in \S~\ref{sec|evolution}, that the average
BH radiative efficiency should be around $\epsilon \lesssim 0.1$, in the assumption that most of the BH mass in the local Universe was accreted during AGN visible phases. In addition, they were the first to point out that, once $\epsilon$ is fixed using Eq.~(\ref{eq|soltan}), the overall shape of the local BHMF can be naturally
reproduced by direct integration of the AGN luminosity over time using
Eq.~(\ref{eq|conteq}), once the appropriate median \lam\ has been adopted.

\begin{figure}[ht!]
\centering
\includegraphics[angle=0,scale=1.0]{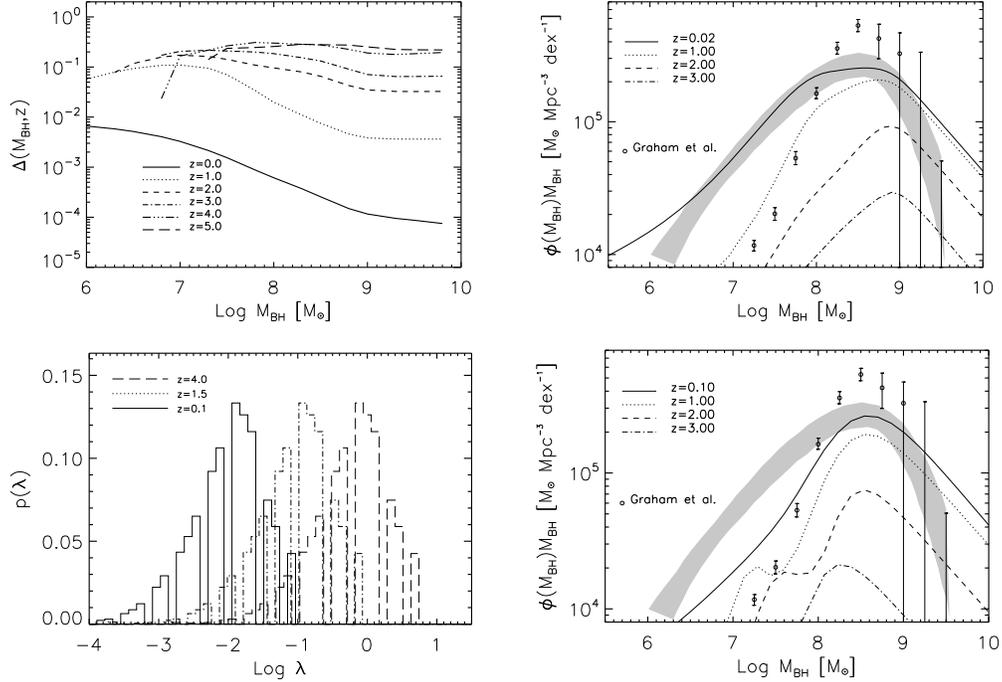}
\caption{\emph{Lower right panel}: predicted duty cycle $\Delta(M_{\rm bh},z)$ on the basis of Eq.~(\ref{eq|conteq})
at different redshifts and BH mass, as labeled. \emph{Upper
right panel}: predicted black hole mass function $\Phi(M_{\rm
bh},z)$ from a model with a delta-function \PL.
The result is shown at different redshifts, from $z=3$ to $z\sim 0$, as labeled,
in the plane $\Phi(M_{\rm bh},z)M_{\rm bh}$, which plots the mass
density per bin of BH mass. The local mass function is shown with a
\emph{grey band} and has a shallower slope than the one predicted by
this model at $z\sim 0$ (\emph{solid} line).
The \emph{open circles} with error bars
show the estimate of the local mass function by Graham et al. (2007).
\emph{Lower left panel}: assumed Gaussian \PL\ distribution
with the peak steadily decreasing with decreasing
redshift. \emph{Lower right panel}: predicted black hole mass function $\Phi(M_{\rm
bh},z)$ from a model with a \PL\ as given in the lower left panel.
Even in this case the result
is shown at different redshifts, as labeled.}
\label{fig|RefModelSWM}
\end{figure}

The upper panels of Figure~\ref{fig|RefModelSWM} show the results of
a reference model which evolves the predicted mass function
assuming an Eddington ratio of $\lambda \sim 0.40$ and $\epsilon=0.065$ (see \citealt{SWM} for further details).
The upper left panel of Figure~\ref{fig|RefModelSWM} plots the duty cycle $\Delta(M_{\rm bh},z)$ as a function of mass for different redshifts, as labeled. The duty cycle for $M_{\rm bh}\sim 10^9\, {\rm M_{\odot}}$
is $\sim 0.2$ at $z\sim 4-5$, falls to 0.03-0.08 at $z=2-3$ when quasar activity
is at its peak, then drops to 0.003 at $z=1$ and $\sim 10^{-4}$ at $z=0$. Below $z\sim 3$, the duty cycle
rises towards low BH masses. This ``downsizing'' evolution, in which high mass BHs
complete their growth early but low mass BHs continue to grow at late times, is required
by the observed luminosity function evolution \citep[e.g.,][]{Ueda03}
in any model with approximately constant \lam.

This model yields overall good agreement with the
average determinations of the local BHMF.
The upper right panel of Figure~\ref{fig|RefModelSWM} plots
$M_{\rm bh}\Phi(M_{\rm bh})$, proportional to the fraction of
BH mass per logarithmic interval of $M_{\rm bh}$, which
allows better visual comparison to the observed local BHMF
and highlights the contribution of each BH mass bin to
the total mass density at each time. The evolution of the mass
function derived from the continuity equation
is shown here at different redshifts from $z=3$ (dot-dashed line)
down to $z=0$ (solid line). Because of the
luminosity-dependent density evolution in the observed luminosity
function, the massive end of the BH mass function builds up
early, and the lower mass regime grows at later redshifts.

For $M_{\rm bh}>10^{8.5}\, {\rm M_{\odot}}$, the mass function is
almost fully in place by $z=1$. At very high masses the model exceeds the observational
estimate, but in this regime the estimate relies on extrapolation of
the scaling relations, and it is sensitive to the assumed intrinsic scatter.
In addition, the high-mass end of the predicted mass function is sensitive to the
bright end of the AGN luminosity function which might
also suffer from some systematic uncertainties (see
\citealt{Fontanot07LF}, \citealt{SWM} and references therein).
The open circles with error bars
show the estimate of the local mass function by \citet{Graham07BHMF}, which
cannot be reproduced even approximately with constant \lam-models.
If this estimate is correct, then the low end of the
luminosity function must be produced mainly by high mass black holes accreting at low \lam, so that the predicted growth of low mass BHs is reduced.

\begin{figure}[ht!]
\centering
\includegraphics[angle=0,scale=1.4]{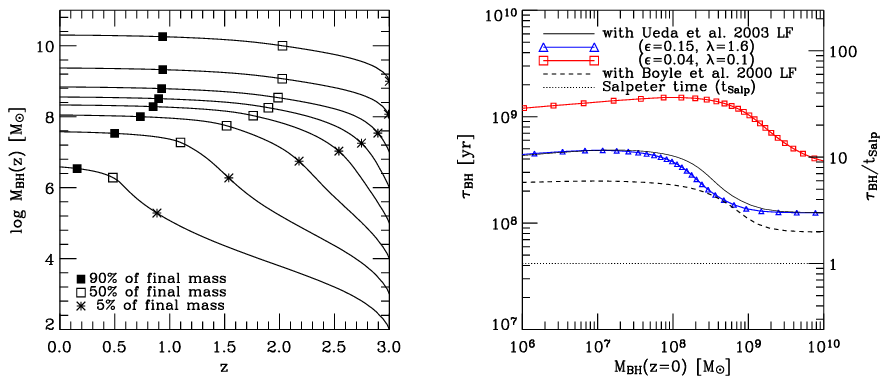}
\caption{\emph{Left panel}: average growth history of BHs with given starting mass at $z=3$
computed using the \citet{Ueda03} luminosity function and $\epsilon = 0.1$, $\lambda = 1$.
The symbols, \emph{filled squares}, \emph{empty squares}, and \emph{stars},
indicate when the black hole reaches 90\%, 50\%, and 5\% of its final mass, respectively. \emph{Right panel}: average mean lifetime of active BHs (AGNs) as a function of the relic BH mass at $z = 0$ computed with $\epsilon = 0.1$, $\lambda = 1$, the
\citet{Ueda03} (\emph{solid} line) and \citet{Boyle00} (\emph{dashed} line) luminosity function. The \emph{dotted} line is the corresponding Salpeter
time. The scale of the \emph{y} axis on the right is the average lifetime in units of the Salpeter time. The lines with the \emph{empty squares} and
\emph{triangles} are the average lifetimes computed from the Ueda et al. (2003) luminosity function with
$\epsilon = 0.04$, $\lambda = 0.1$, and $\epsilon = 0.15$,
$\lambda = 1.6$, respectively. Figures taken from \citet{Marconi04}.}
\label{fig|Marconi}
\end{figure}

Assuming that all BHs at all times radiate at
a constant Eddington ratio is obviously an oversimplification. Some
empirical evidence for some sharp Eddington ratio distribution
$p(\lambda,z)$ has been found  by \citet{Kollmeier06} who claim
that most AGNs of different BH mass in the AGES survey at $0.5\lesssim z \lesssim 4$ all share a similar Gaussian-shaped $p(\lambda)$, constantly peaked around $\lambda\sim 0.25$ and with a dispersion of $\sim 0.3$ dex. On the other hand, at least at $z\lesssim 1$,
AGN Eddington ratio distributions have a wide distribution, with a large fraction of Seyferts radiating
at $\lambda < 0.1$ \citep[e.g.,][]{Hickox09,Kauffmann09}. Some studies also show evidence for
strong redshift evolution and milder mass-dependence of the typical Eddington ratio \citep[e.g.,][]{Heckman04,Veste04,MD04,Constantin06,Dasyra07,NetzerT07,Rovilos07,Shen08bias}, though systematic uncertainties in the
reverberation mapping techniques and extrapolations of empirical
virial relations \citep[e.g.,][]{Kaspi00,Bentz06}
make it hard to draw firm conclusions.

The lower left panel of Figure~\ref{fig|RefModelSWM} shows
an input $p(\lambda,z)$ with negligible mass-dependence
but a strong redshift evolution (as labeled in the panel)
required to be simultaneously consistent
with the high-$\lambda$ observed at $z>1$ by \citet{Kollmeier06}, and the
low $\lambda\sim 0.01$ observed in SDSS by, e.g., \citet{Kauffmann09}. At each redshift
a significant spread in $p(\lambda,z)$ consistent
with the observed one is also implemented.
The predicted BH mass function $\Phi(M_{\rm bh},z)$ implied by this model
is shown in the lower right panel of Figure~\ref{fig|RefModelSWM},
at different
redshift from $z=3$ to $z\sim 0$ as labeled, plotted in the
$\Phi(M_{\rm bh},z)M_{\rm bh}$ plane. From the comparison
with the local BHMF, it is clear that the latter model
still predicts a similar break for the $z=0$ BH mass function as in
constant $\lambda$-models, because the average $\lambda$ is still about $0.3-0.5$.
However, much less numerous low mass BHs are produced when the median \lam\ strongly decreases with redshift, because more of the low-$z$ emissivity is mapped into more massive
BHs ($L\propto \lambda M_{\rm bh}$), preventing the growth of the less massive ones. It is therefore clear that an extremely accurate
estimate of the local BHMF is fundamental to discern among different BH accretion models (see \S~\ref{subsec|understandingtheAGNLF}). It is also possible to show that broadening the input \PL\ distribution does not affect the downsizing of BHs at low redshifts,
but produces an \emph{upsizing} at higher redshifts (see also \citealt{MerloniHeinz08}),
with the fraction of more massive active BHs significantly higher than low
mass ones.

Other interesting quantities which can be directly derived
from the continuity equation formalism developed above are the mean growth of BHs
of a given mass at $z=0$ and the average visible lifetime of BHs.
The former quantity can be directly drawn from direct integration
on the mean accretion rate
\begin{equation}
M_{\rm bh}(z)=\int_{\infty}^{z}\langle \dot{M}_{\rm bh}(z') \rangle \frac{dt}{dz'}dz'\,.
\label{eq|Mbht}
\end{equation}
The result is shown in the left panel of Figure~\ref{fig|Marconi} as derived
from \citet{Marconi04}, which shows
the average growth history of BHs from $z=3$
computed using the \citet{Ueda03} luminosity function and $\epsilon = 0.1$, $\lambda = 1$.
The symbols, filled squares, empty squares, and stars,
indicate when the BH reaches 90\%, 50\%, and 5\% of its final mass, respectively.
Here it is apparent the notion of downsizing, where the most massive
BHs acquire most of their mass at $z>1$, at variance with lower mass BHs.

The corresponding time for which a BH of a given mass at $z=0$
was active in the past can be derived by direct integration of the duty cycle over time, as
\begin{equation}
t_{\rm vis}[M_{\rm bh}(z=0)]=\int_{\infty}^0 \Delta(M_{\rm bh},t) dt\, .
\label{eq|tvis}
\end{equation}
The result, again taken from \citet{Marconi04}, is shown in the right panel of Figure~\ref{fig|Marconi}
for a model with $\epsilon = 0.1$ and $\lambda = 1$, using both the X-ray and optical luminosity
functions of
\citet{Ueda03} (solid line) and \citet{Boyle00} (dashed line), respectively. Also the outputs of other models with slightly different accretion parameters are plotted in the same Figure for comparison. It is clear that lower mass BHs have longer visibility timescales than more massive BHs. Note, however, that the latter timescale strongly depend on the epoch of ``first appearance'' in which the BH of that final mass starts shining in the luminosity function, and therefore cannot be precise estimates of BH lifetimes, as discussed by several Authors \citep[e.g.,][]{Hosokawa02,YT02,Shankar04,YuLu04,HopkinsHernquist08a}.


\section{Black Hole mergers}
\label{subsec|merging}

So far we have assumed that BHs mainly grow through
accretion. However, observations show, and hierarchical galaxy formation models predict, that a significant fraction of galaxies experience mergers with comparably massive galaxies during their
lifetime. At least some of these galaxy mergers are likely to be accompanied by mergers of
the central BHs that they contain.
In fact, as the galaxies merge, the BHs should sink toward the center of the new galaxy via dynamical friction where they form a binary, which
can continue its decay by transferring angular momentum to the stars intersecting its orbit.
If the binary's separation decreases to the point where the emission of gravitational waves becomes efficient at carrying away the last remaining angular momentum, the BHs coalesce rapidly. However, the mechanisms that shrink the orbits of the BHs with their small cross-sections
to such small scales of the order of a $\sim 1$ parsec,
are not fully understood (see \citealt{Merritt05} for a review).

In principle, including mergers in the overall evolution of the BH population,
requires inserting a mass and redshift-dependent source term in Eq.~(\ref{eq|conteq}).
During the evolution, mergers will then redistribute mass within the BH population,
thus varying at any given epoch the predicted mass distribution with respect
to a pure accretion model.
Moreover, mergers should not change the integrated mass density, so they do not affect
the integrated \citet{Soltan} argument discussed in the previous sections;
in principle, gravitational radiation
during mergers can \emph{reduce} the integrated mass density, but it should
not be a dominant effect (see \citealt{YT02}).

However, at present it is still difficult
predicting precisely the BH mass function implied by gas accretion
and mergers. On the one side,
the actual rate of BH merging is still highly uncertain;
on the other side, the indirect important effect that mergers might have on the
mass-dependence of the radiative efficiency is still unknown.
In fact, there is a tight connection between radiative efficiency and angular momentum
of the BH. Most of the energy is radiated close to the innermost stable orbit,
which gets closer to the central rotation axis if the BH is rapidly spinning
(see, e.g., \citealt{Bardeen72} for the formalism).
Therefore a high spin is paralleled by a high radiative efficiency.
BHs are believed to spin up or spin down after mergers
and/or gas accretion, although the details of such processes
are too poorly understood to allow detailed modeling. The canonical value of the radiative efficiency competing to a non-rotating BH is $\epsilon\sim 0.06$, increasing up to $\epsilon \sim 30\%$ for a rapidly spinning BH in equilibrium \citep{Thorne74}.
The value of the radiative efficiency $\epsilon \lesssim 0.1$ inferred from the \citet{Soltan} argument discussed in \S~\ref{sec|evolution} (see Figure~\ref{fig|Soltan}), appear to indicate that on average BHs are not extreme rotators, although the caveats discussed
in \S~\ref{sec|evolution} might bias this conclusion.

\citet{HughesBlandford} studied the remnants of merging BHs of different
sizes, spins and orbital parameters. They concluded that BHs are
generally spun down by mergers and only for a very narrow range of
orbital parameters merging is effective in rapidly spinning up the
holes. \citet{Gammie04} carried out numerical simulations
confirming the \citet{HughesBlandford} result that a BH is
usually spun down by frequent minor mergers, while major
mergers usually spin up the BHs.

\citet{Volonteri05} studied the expected distribution of BH spins and its evolution with cosmic time in the context of hierarchical galaxy formation theories. They followed the merger history of dark matter halos as extracted from Monte Carlo realizations based on the Extended
Press-Schechter formalism, and followed the growth of BHs via gas accretion
and mergers. They concluded that mergers do not lead to a systematic spin up or spin down of BHs with time, but actually the spin distribution retains memory of its initial conditions. In their models, accretion is the main driver for efficiently spinning BHs up, and they find that if accretion takes place from a thin disk, then up to 70\% of all BHs can be maximally rotating and have radiative efficiencies $\gtrsim 12\%$.
The \citet{Volonteri05} conclusion was based on the
assumption that the BH spin aligns very rapidly with the accretion flow, although they also note that a randomly--oriented sequence of accretion events is less efficient in spinning the BH up.
\citet{KingPringle}, strengthen the latter point, discussing
through analytical methods how a randomly oriented
sequence of accretion episodes can effectively keep the spin
low, rather than aligning the hole very quickly and causing systematic spin up.
\citet{BertiVolonteri}
used results from numerical relativity simulations of BH mergers, coupled to the cosmological evolution of BH spin, to infer the evolution of the spin distribution. They conclude that mergers are very unlikely to yield substantial spins, unless alignment of the spins of the merging holes with the orbital angular momentum is very effective.

\citet{SWM} have illustrated the potential impact
of mergers on the predicted BH mass function
using a simple, yet useful mathematical model
that assumes constant probability
of equal mass mergers per Hubble time,
similar to the models of \citet{SteedWeinberg}.
Following the latest results discussed above on the relatively mild
effect of mergers on the spin distribution,
\citet{SWM} neglected the variations of radiative efficiency
caused by the impact of mergers on the BH
spin distribution, and also secondary effects such as ejection of BHs
by gravitational radiation or three-body interactions \citep[e.g.,][]{HughesBlandford,Gammie04,Islam04,Yoo04,Volonteri05,Merritt05,Yoo07,Sesana09}.

\citet{SWM} assumed that a BH of mass \mbh\ has a probability
$P_{\rm merg}$ of merging with an equal mass black hole in the
Hubble time $t_H(z)$ (age of the Universe at redshift $z$).
Therefore the fraction $F$ of black holes that merge in a timestep
$\Delta t$ is given by
\begin{equation}
F=P_{\rm merg}\times\frac{\Delta t}{t_H(z)}\, .
    \label{eq|mergF}
\end{equation}
At each time $t_1$, these Authors first advance the mass function to time
$t_2=t_1+\Delta t$ with accretion only, then add to each bin of the mass function
an increment (which may be positive or negative)
\begin{equation}
\Delta \Phi(M_{\rm bh},t_2)=\frac{F\times \Phi
\left(\frac{M_{\rm bh}}{2},t_2\right)}{2}-F\times
\Phi(M_{\rm bh},t_2)\, ,
    \label{eq|merging}
\end{equation}
where the second term represents BHs lost from the bin by merging
and $\Phi \left(\frac{M_{\rm bh}}{2},t_2\right)$ is calculated by
interpolation.

\begin{figure}
\includegraphics[angle=00,scale=0.8]{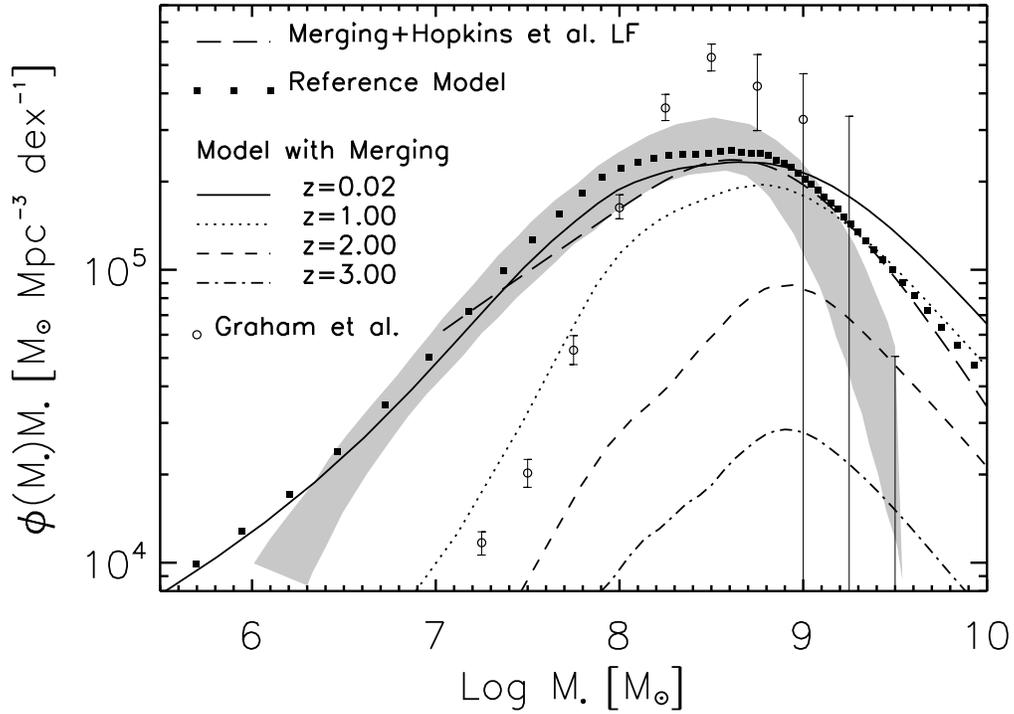}
\caption{Evolution of the black hole mass function in a model with black
hole mergers. Accretion-driven growth is computed assuming $\epsilon=0.065$,
$\lambda=0.40$, and each black hole has a probability $P_{\rm merg}=0.5$
of merging with another black hole of equal mass per Hubble time $t_H(z)$.
Squares show the $z=0$ predictions of the reference model (same accretion
parameters, no merging), and the long-dashed line shows the $z=0$ mass function
for a merger model with the \citet{Hop07} luminosity function and accretion parameters $\epsilon=0.09$,
$\lambda\sim 1$.} \label{fig|merging}
\end{figure}

Figure~\ref{fig|merging} shows the evolution of the BH mass function resulting from a model
with $P_{\rm merg}=0.5$,
$\epsilon=0.065$ and $\lambda\sim 0.4$. The net effect of
merging is to slightly lower the abundance of BHs below the peak
of the mass function and to significantly increase the abundance of very massive
BHs, as expected
(see also \citealt{Malbon07}, \citealt{Yoo07}).
Since a non-merger model already produces an excess of massive BHs
relative to the local BHMF (see Figure~\ref{fig|RefModelSWM}), adding mergers only makes
the match to observations worse. However, the impact of mergers is evident
mainly for $M_{\rm bh}>10^9\, {\rm M_{\odot}}$, where the BHMF
estimates are most sensitive to the adopted scatter in the black hole-host scaling
relations and to the extrapolations of these relations up to the most
luminous galaxies. Note that the value $P_{\rm merg}=0.5$ is an upper limit
for the merger rate of massive galaxies (see, e.g., \citealt{Bundy09} and references therein),
and it was chosen to provide an upper
limit on the actual impact of mergers on the BH population.


%

The impact of mergers on
the predicted BH mass function is likely to be small compared to remaining
uncertainties in accretion-driven growth, except perhaps for the
rare, high mass BHs. The most interesting impact of BH mergers may arise indirectly,
through their effect on BH spins and thus on radiative efficiencies, as discussed above \citep[e.g.,][]{Volonteri05}.
Assuming, for example, the radiative efficiency to increase with increasing
BH mass, might be a viable solution to accommodate a sequence of numerous
mergers with the extremely steep fall-off of the local BHMF at the high-mass end (see also \citealt{CaoLi08}). Potential
evidences for some mass and/or redshift dependence of the radiative efficiency may already
be available (see \S~\ref{sec|clustering}).

\begin{figure}
\centering
\includegraphics[angle=00,scale=0.7]{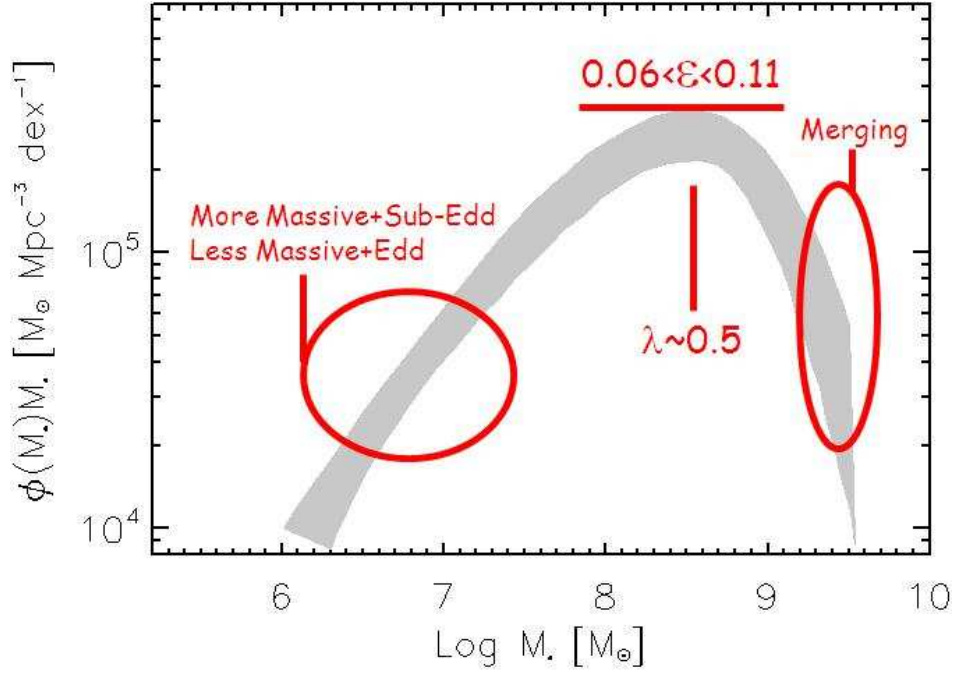}
\caption{Summary of the main constraints which
can be derived by comparing the predictions of accretion plus mergers models for black hole evolution
with the local black hole mass function (grey area). The overall height under the curve
constrains the radiative efficiency, while the peak of the mass distribution
is able to constrain the mean Eddington ratio. Matching the low end of the mass distributions can instead shed light on the alternative between models with low-mass BHs growing at a significant
fraction of the Eddington limit, and models marked by higher-mass BHs
growing at sub-Eddington regimes. The high-mass end, coupled with some information
on the mass-dependence of the radiative efficiency, should instead provide useful constraints
on the impact of mergers on the overall BH evolution.} \label{fig|summary}
\end{figure}

Figure~\ref{fig|summary} illustrates my summary of the main constraints which
can be derived by comparing the predictions of accretion plus mergers models for BH evolution
with the local BHMF. The overall BH mass density
constrains the radiative efficiency as first pointed out by \citet{Soltan}, while the peak of the mass distribution
is able to constrain the \emph{mean} Eddington ratio (\citealt{Salucci99}). Matching the low end of the mass distributions can instead shed light on the alternative between models with low-mass BHs growing at a significant
fraction of the Eddington limit, and models marked by higher-mass BHs
growing at sub-Eddington regimes (see \S~\ref{subsec|BHMFevol}). Finally, the high-mass end, coupled with some information
on the mass-dependence of the radiative efficiency, should provide useful constraints
on the impact of mergers on the overall BH evolution.

\section{AGN CLUSTERING PROVIDES INDEPENDENT CONSTRAINTS ON BLACK HOLE EVOLUTION}
\label{sec|clustering}

In the previous sections it has been
shown that continuity equation arguments
based on evolving the BH population linking
BH growth to the observed AGN bolometric luminosity function
yields interesting constraints when compared to the local BHMF.
However, this technique can only constrain \emph{mean} quantities.
For example, it cannot distinguish between a model with a constant
$\lambda=0.4$ at all redshifts, or a model with Super-Eddington accretion
($\lambda>1$) at higher redshifts, steadily decreasing to
sub-Eddington regimes at lower-$z$ (lower panels of Figure~\ref{fig|RefModelSWM}).
On the other hand, there
are no assumption-free direct observations of the radiative
efficiency at high redshifts \citep[e.g.,][and references therein]{Rafiee09,Wang09},
and a constant as well as a redshift-dependent radiative
efficiency can still satisfy the \citet{Soltan} argument. Breaking
such degeneracies would require knowledge of the BH mass
function at higher redshifts, which is still not feasible given that
it is still not clear whether the local relations between BHs and
galaxies still hold at higher redshifts \citep[e.g.,][and references therein]{ShankarMsigma}.

Quasar clustering provides additional, independent constraints on
the main parameters regulating BH evolution \citep[e.g.,][]{MoWhite,Haehnelt98}
and their
possible evolution with redshift and/or mass, thus helping in discriminating among the
successful models. As first outlined by \citet{MW01}
and Haiman \& Hui (2001; see also \citealt{Lidz06,Hop08clust,ShankarCrocce,White08,Bonoli09,Marulli09,Thacker09}),
the clustering is in fact a
direct measure of the masses, and therefore number densities, of the
halos hosting the quasars. In turn, the ratio between the quasar
luminosity function and the halo mass function provides information
on the duty cycle, i.e., the fraction of halos which shine as
quasars at a given luminosity, BH mass, or Eddington ratio.

Many groups have now been able to measure
the clustering of AGNs at different luminosities, bands, scales, and redshifts
\citep[e.g.,][and references therein]{Overzier03,Grazian04,Croom05,Constantin06,Hennawi06,Porciani04,Porciani06,daAngela08,Myers07a,Myers07b,Padmanabhan08,Shen07,Shen09,Coil07,Coil09,Miyaji07,Hickox09,Ross09}. AGN clustering is usually expressed in terms of the bias, i.e., the square root of the ratio
between the matter and the measured two-point correlation functions.
It is beyond the scope of this Review to discuss
and compare all the above mentioned empirical results. In the following
we will just focus on the semi-empirical methods which have been adopted
in the literature to actually use such measurements for deriving firm
constraints on BH evolution.


The classical modeling of quasar clustering by Haiman \& Hui (2001)
and Martini \& Weinberg (2001) assumes a mean value for the duty
cycle and derives the relation between quasar luminosity and host
halo mass by monotonically matching their cumulative distribution
functions. Formally, this concept leads to
\begin{eqnarray}
\int_{x_{\rm min}}^{\infty} \Phi(x,z)\, dx= \int_{0}^{\infty}dy \, \Delta_h(y,z)\Phi_h(y,z)\times \nonumber\\
\frac{1}{2}{\rm erfc}\left[\ln
\left(\frac{10^{y_{\rm min}(x_{\rm min})}}{10^y}\right)\frac{1}{\sqrt{2}\ln(10)\Sigma}\right]\, ;
\label{eq|cummatching}
\end{eqnarray}
here $x=\log L$ ($L$ being the quasar luminosity), and $y=\log M_h$ ($M_h$ being the halo mass).
The quantity $\Phi_h(y,z)$ is the comoving number density of halos,
in units of ${\rm Mpc^{-3}}\, {\rm dex}^{-1}$ \citep[e.g.,][]{ShethMF},
while $\Phi(x,z)$ is the luminosity function of quasars in the same units of ${\rm Mpc^{-3}}\, {\rm dex}^{-1}$.
The quantity $\Delta_h(y,z)$ in
Eq.~(\ref{eq|cummatching}) is the duty cycle, i.e., the fraction of
halos which host quasars shining above a minimum luminosity
$x_{\rm min}=\log L_{\rm min}$ at redshift $z$.
Eq.~(\ref{eq|cummatching}) also takes into account a scatter of
$\Sigma$ (in dex) between quasar luminosity and halo mass.

So the clustering clearly provides an independent way
to constrain duty cycle $\Delta_h(y,z)$ of active halos and BHs, which can be directly
compared with the $\Delta(M_{\rm bh},z)$ predicted from continuity equation arguments
(see Eq.~[\ref{eq|P0general}]).
Either method can be adopted independently to work
simultaneous constraints on the radiative efficiency, Eddington ratio distributions,
and the underlying
BH mass distribution at a given redshift; this ultimately provides
in ways independent of specific models, powerful tools to
delineate a comprehensive picture of BH evolution.
Also, these determinations of the duty cycles can be further constrained
by comparing with direct estimates of the fraction of active galaxies
in complete samples above a certain luminosity \citep[e.g.,][]{Kauffmann03}.

At each redshift Eq.~(\ref{eq|cummatching}) defines the minimum
halo mass $y_{\rm min}$ corresponding to the minimum luminosity in the
sample $x_{\rm min}$. The mean bias associated to the same
subsample at redshift $z$ is then
\begin{equation}
\langle b\rangle =\frac{\int_{y_{\rm min}}^{\infty}\, dy\, \Phi_h(y,z)b(y,z)}{\int_{y_{\rm min}}^{\infty}\,
dy\, \Phi_h(y,z)}\, .
    \label{eq|beff}
\end{equation}

\begin{figure}
\centering
\includegraphics[angle=00,scale=1.5]{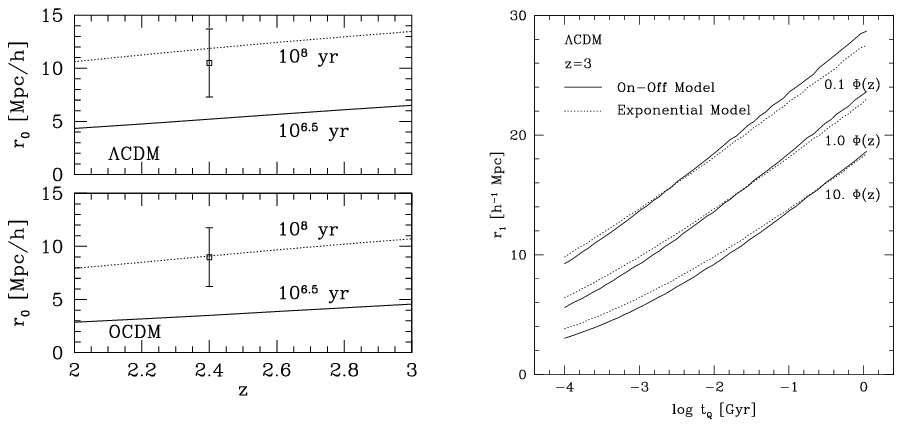}
\caption{\emph{Left panel}: correlation length predicted for two different quasar lifetimes, as labeled.
Figure taken from \citet{HH01}. \emph{Right panel}: clustering length vs quasar lifetime $t_Q$ and for different values of the quasar number density $\Phi$, as labeled. Figure taken from \citet{MW01}. The main point in both panels is that an
higher quasar lifetime (i.e., an higher duty cycle) implies
the quasars to be hosted in less numerous, more massive and more biased halos.} \label{fig|clustering}
\end{figure}

\citet{MW01} and \citet{HH01} applied Eq.~(\ref{eq|cummatching}) showing that quasar clustering
measurements can substantially narrow down the range of possible duty cycles.
In fact, if quasars are long-lived they are rare
phenomena that are highly biased with respect to the underlying dark matter; whilst if short
lived, they reside in more typical halos that are less strongly clustered. Figure~\ref{fig|clustering}
shows examples of their calculations in which the duty
cycles are actually expressed in terms of the quasar optical lifetime $t_Q$, which is directly
proportional to the duty cycle $\Delta_h$, for given approximations as to the lifetime of the host halos. Both
plots show that the clustering strength increases for increasing quasar lifetime $t_Q$ and, at fixed
$t_Q$, increases for decreasing quasar number density $\Phi$, as labeled.
This behavior is easily understood given that
longer quasar lifetimes (i.e., an higher duty cycle), and/or lower quasar abundances, imply
quasars to be hosted in less numerous, more massive, and more biased halos.

\begin{figure}
\center
\includegraphics[angle=00,scale=0.6]{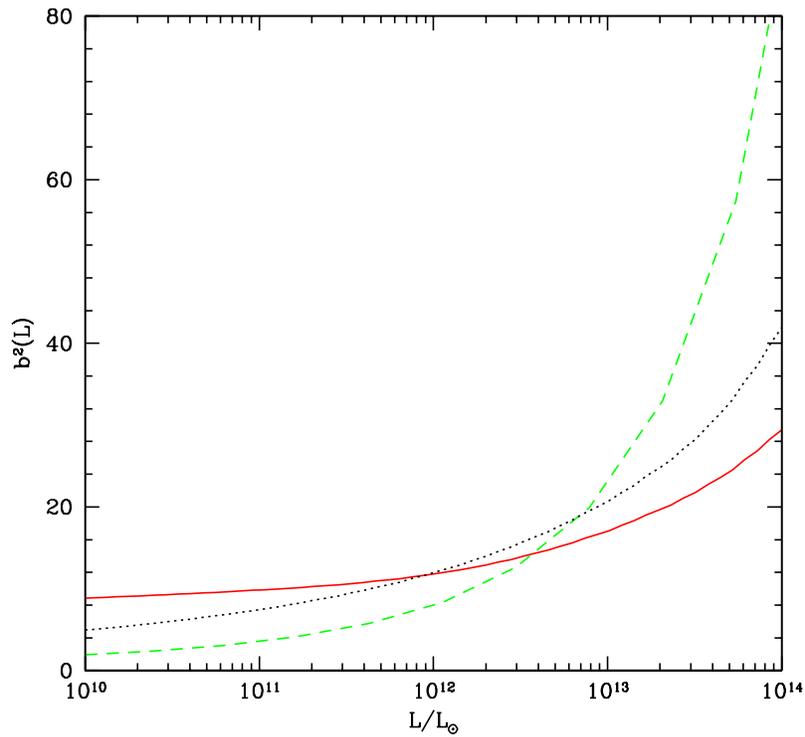}
\caption{Bias-squared of quasars as a function of their luminosity.
The \emph{solid} and \emph{dotted} lines show the luminosity dependence of quasar clustering in a model with a non-linear light curve as predicted from numerical simulations, while the \emph{dashed} line refers to a strictly monotonic
model between quasar luminosity and halo mass. Figure taken from \citet{Lidz06}.} \label{fig|Lidz}
\end{figure}

Eq.~(\ref{eq|beff}) can be generalized to compute the bias as a function
of quasar luminosity. Assuming that quasar luminosity closely tracks halo
mass, one would naively expect an increasing clustering strength for more
luminous quasars. While this seems to indeed be the case for
high-redshift quasars \citet{Francke08,Shen09}, lower redshift
surveys find evidence instead for a much flatter quasar bias as a function
of luminosity. \citet{Lidz06} investigated the luminosity dependence of quasar clustering in more detail and concluded that by adopting more complicated, non-monotonic quasar light curves, such as the ones predicted from numerical simulations \citet{Hopkins06}, a better match to the low-$z$ data can be found.
This is because in this picture the Eddington ratio distribution
at fixed BH mass is much broader, thus mapping both fainter and brighter quasars
into similar bins of BH/halo mass, and therefore flattening the bias
as a function of luminosity. The result of their model is shown in Figure~\ref{fig|Lidz} (dotted and solid lines), compared to a model (dashed line) which maps quasars
into halos with a nearly one-to-one relation, and thus predicts a much stronger rise
of the bias with increasing luminosity.

\begin{figure}
\center
\includegraphics[angle=00,scale=0.7]{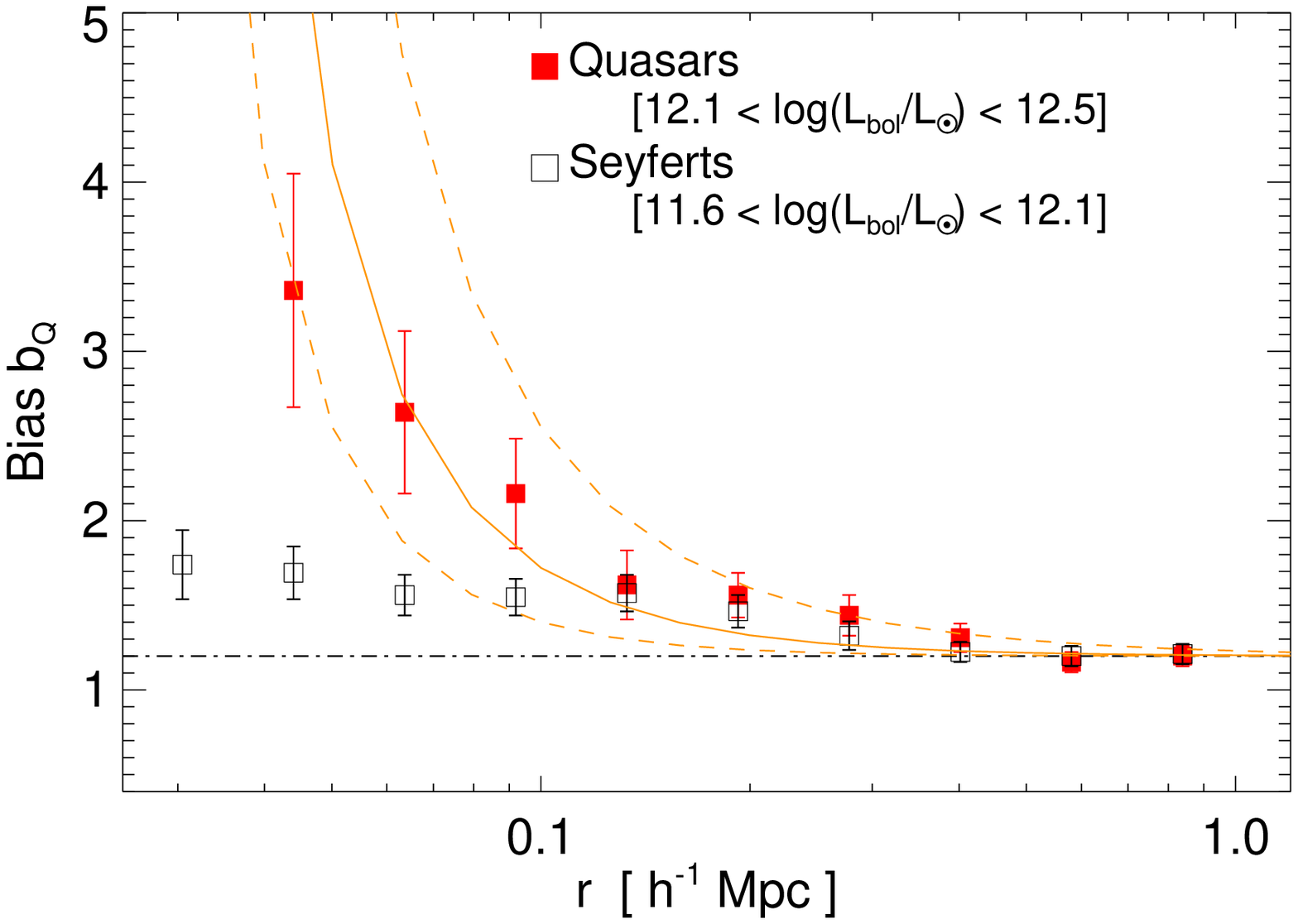}
\caption{Excess small-scale clustering of quasars and Seyferts.
The data are from \cite{Serber06} for optical quasars (red filled squares)
and Seyferts (black open squares).
Plotted is the predicted bias as a function of scale radius determined from the correlation function at each radius relative to that at large scales
(i.e., dividing out the best-fit large-scale dark matter correlation function). The \emph{dotted–dashed} line shows a constant bias as a function of scale (i.e., no excess of
small-scale densities, expected if the galaxies are a random subpopulation with no preference for small groups), the \emph{solid} (\emph{dashed}) orange line shows the predicted
excess on small scales expected for recent merger populations from cosmological models and simulations
(reflecting a preference for small group environments). Figure taken from \citet{HopkinsHernquist09b}.} \label{fig|smallscales}
\end{figure}

Other important constraints can also be derived
by studying the scale-dependence of the quasar bias.
If luminous quasars are predominantly triggered by merging/interaction
events, then their clustering strength at small scales should
be enhanced with respect to a random non merger population
with similar masses and large-scale clustering properties.
Indeed, such a small-scale excess has been seen, in both low-redshift \citet{Serber06}
and high-redshift \citet{Hennawi06,Myers07b} quasar populations. Interestingly,
however, \citet{Serber06} see no such excess in the low-
redshift Seyfert population, as shown in Figure~\ref{fig|smallscales},
which plots the excess small-scale clustering of quasars and Seyferts
with data taken from \citet{Serber06} for optical quasars (red filled squares)
and Seyferts (black open squares).
The lines show the predicted predicted bias as a function of scale radius determined from the correlation function at each radius relative to that at large scales, as predicted
from the evolutionary models of \citet{HopkinsHernquist09b}.
Interestingly, \citet{Li08} using SDSS data showed that while specific star formation rates of galaxies are higher if they have close neighbors, close neighbors are not associated with any similar enhancement of nuclear activity. This leads to conclude that star formation induced by a close companion and star formation associated with BH accretion are distinct events. They also suggest that these events might still be part of the same physical process, for example a merger, provided they are separated in time. These empirical results, as well as those by \citet{Serber06}, support a scenario in which lower luminosity, often lower mass BHs, as usually associated with Seyfert galaxies, might be more likely triggered by in-situ phenomena, at variance with the violent dynamical events
which are believed to be at the origin of the high-$z$, luminous quasars (e.g., \citealt{Sanders88}, see also \citealt{Kawakatu03} and \S~\ref{sec|models}).

Particularly interesting constraints
can be derived from the recent $z>3$ clustering
measurements of luminous quasars in SDSS by \citet{Shen07}.
\citet{White08} have applied Eqs.~(\ref{eq|cummatching}) and (\ref{eq|beff})
to the data, concluding that the strong
clustering measured at $z\sim 4$ can be understood only if quasar
duty cycles are high and the intrinsic scatter in the
luminosity-halo relation is small.
\citet{ShankarCrocce} took a further
step by jointly considering the evolution of the BH-halo
relation and the BH mass function, as constrained by the
observed AGN luminosity function and clustering. They examined
constraints on the host halos, duty cycles, radiative efficiencies,
and mean Eddington ratios of massive BHs at $z>3$, as imposed
by the clustering measurements of \citet{Shen07} and by a variety of
measurements of the quasar luminosity function at $3 \le z \le 6$
\citep[e.g.,][and references therein]{Hop07,ShankarMathur,Fontanot07LF,SWM}.
They linked BH growth to halo growth via a relation between BH mass and halo
virial velocity $V_{\rm vir}$.
They found that the strong clustering of AGNs observed at
$z=3$ and, especially, at $z=4$ implies that massive BHs
reside in rare, massive dark matter halos. In turn, reproducing the observed
luminosity function as implied by the growth of dark matter halos,
requires high efficiency $\epsilon$ and/or
low Eddington ratio $\lambda$, with a lower limit (based on
$2\sigma$ agreement with the measured $z=4$ correlation length)
$\epsilon\gtrsim 0.7\lambda/(1+0.7\lambda)$, implying $\epsilon \gtrsim
0.17$ for $\lambda > 0.25$. The rapid drop in the abundance of the massive and
rare host halos at $z>7$ also implies a proportionally rapid decline in
the number density of luminous quasars, much stronger than simple
extrapolations of the $z=3-6$ luminosity function would predict.

\begin{figure}
\center
\includegraphics[angle=00,scale=1.1]{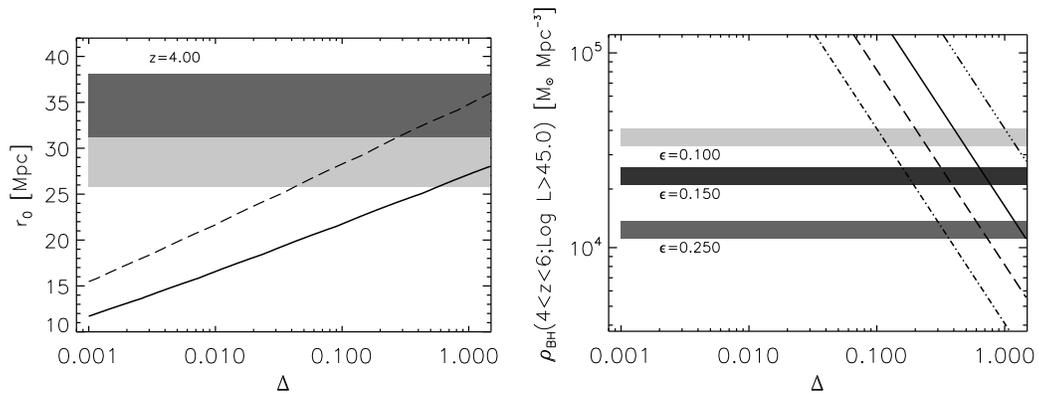}
\caption{\emph{Left panel}: predicted $z=4.0$ clustering
correlation length $r_0$ computed above the minimum survey
sensitivity as a function of duty cycle, having adopted the luminosity
function derived from the reference model by \citet{ShankarCrocce}.
The \emph{solid} and
\emph{long-dashed} lines refer to the $r_0$ implied by using
different bias formulae (see \citealt{ShankarCrocce}).
Dark and light shaded bands show the $1\sigma$
range of the \citet{Shen07} measurements at these redshifts.
Maximal values of the
duty cycle predict a clustering strength only marginally consistent
with the data at $z=4$. \emph{Right panel}: \emph{horizontal stripes}
show the integrated black hole mass
density above $L=10^{45}\, {\rm erg\, s^{-1}}$ and $4.0<z<6$ derived
from the AGN luminosity function for three different
values of the radiative efficiency, as labeled; \emph{triple dot-
dashed}, \emph{solid}, \emph{long-dashed}, and \emph{dot-dashed}
lines indicate the black hole mass density implied by the AGN luminosity
function at $z=4.0$ integrated over mass assuming a mean Eddington
ratio $\lambda=0.1,0.25,0.5,1$, respectively, as a function of duty
cycle $\Delta$. High duty cycles
require high radiative efficiencies or low Eddington ratios to
reconcile the cumulative AGN emissivity with the black hole mass
density in rare halos (see text). Figures taken from \citet{ShankarCrocce}.} \label{fig|r0zDuty}
\end{figure}

Figure~\ref{fig|r0zDuty} shows a compact summary of their main result.
The left panel shows that only assuming maximal values of the
duty cycle $\Delta$ the predicted clustering strength (solid and dashed lines)
can be, at least marginally, consistent with the data at $z=4$ (grey bands).
The right panel shows with horizontal stripes
the integrated BH mass
density above $L=10^{45}\, {\rm erg\, s^{-1}}$ and $4.0<z<6$ derived
from the AGN luminosity function for three different
values of the radiative efficiency, as labeled. The triple dot-dashed, solid, long-dashed, and dot-dashed
lines indicate the BH mass density implied by the AGN luminosity
function at $z=4.0$ integrated over mass assuming a mean Eddington
ratio $\lambda=0.1,0.25,0.5,1$, respectively, as a function of duty
cycle $\Delta$. High duty cycles imply quasars to reside in less numerous, more massive and more
biased halos. Therefore reproducing the high emissivity
of luminous quasars
requires BHs accreting at a significant fraction of the Eddington luminosity
($\lambda \gtrsim 0.25$) to radiate at \emph{high} radiative efficiencies
($\epsilon\gtrsim 0.15$) to match the AGN luminosity function and
its evolution with redshift.

\citet{WL09} also suggested that
the strong $z=4$ clustering could be explained by tying quasar
activity to recently merged halos, which might have stronger bias
(see also \citealt{Wechsler06,FK06,Wetzel09}).
However, halos with substantial ``excess'' bias might be too rare
to satisfy duty cycle constraints, and/or the amount of extra bias might be too modest
to make any significant difference in the models described by \citet{ShankarCrocce}.
Only numerical simulations will be able to provide significant answers
in this direction.

\section{AN OVERVIEW ON COEVOLUTION MODELS OF BLACK HOLES AND GALAXIES}
\label{sec|models}

\subsection{GALAXY DICHOTOMY AND AGN FEEDBACK}
\label{subsec|AGNfeedback}

Galaxies can be classified into two families, which reflect distinct
morphological, chemical and evolutionary features. Following, e.g.,
\citet{Kauffmann03}, \citet{Heckman04}, and \citet{DekelBirnboim},
here are summarized the main pieces of observational evidence
supporting this bimodal classification. There is a gap at
a color of $u-r \sim$ 2, where galaxies naturally divide into a blue
and a red sequence. The former are on average younger, less massive
and disky ({\it late-type} galaxies), while the latter are more
massive older and boxy, spheroidal ({\it early-type} galaxies).
Blue, late-type galaxies dominate the mass function below the knee,
$L^{\star}$, of the Schechter function while red, early-type
galaxies take over above $L^{\star}$.

There is a remarkable correlation between the colors and
luminosities of elliptical galaxies, with brighter galaxies being
redder. Values of the elemental ratios [$\alpha$/Fe], in particular
for Mg, have been found to increase with galactic mass
\citep[e.g.,][]{Terlevich02,Bernardi03}, implying that the
present rate of Supernovae (SN) I is ten times higher than the rate
of SNII. The latter is close to being a tracer of the instantaneous
star formation rate. These observations thereby constitute
significant evidence that massive galaxies formed most of their
stars at $z\geq 2$--3 in a relatively short burst.
Direct evidence that massive galaxies with $M_{\rm star} \ge 10^{11} M_{\odot}$
were in place at $z \ge 2$, is provided by recent $K$-band and
submillimeter surveys \citep[e.g.,][]{Cimatti02,Greve05}. The
fundamental plane relation \citep[e.g.,][and references therein]{DDFP,Treu04,ShankarBernardi} observed up to
$z \sim 1$, is also consistent with massive spheroids being old and
quiescent.

According to the standard cosmological paradigm of structure
formation and evolution, dark matter halos have grown
hierarchically, through the merging together of smaller units into
ever larger systems. In this scenario, galaxies form inside this
hierarchically growing system of halos. However, hierarchical
models do not predict the correct abundances of galaxies at high and
low redshifts \citep[e.g.,][]{Fontana06,MarchesiniVandokkum}
and, at variance with observations, they also tend to
generate negative gradients for the $[\alpha/Fe]$ abundance with
respect to stellar mass, as more $Fe$ is formed in more massive
systems which complete their formation at later times in pure
hierarchical models \citep[e.g.,][]{Thomas99}.

Therefore in recent years theoretical models of galaxy formation have inserted
AGN feedback as a fundamental ingredient for successfully matching several
pieces of independent data, from the stellar mass function to
chemical abundances \citep[e.g.,][]{WL03,Granato04,Scannapieco04,DiMatteo05,Sazonov05,Vittorini05, Bower06,Cattaneo06,Croton06,DeLucia06,Hopkins06,Lapi06,Shankar06,Malbon07,Menci08,Benson07,Monaco07,Cook09}.
However, clear and statistically significant evidence for AGN
feedback in galaxies is still lacking, despite some significant
studies \citep[e.g.,][]{Chartas03,KingPounds,Schawinski07,DAIBAL,Shankar08BAL,Shankar08Cav}.

AGN feedback consists of injection of energy and momentum acting on
the gas and dust of the host galaxy. Its implementation in
theoretical models differs significantly between different research groups;
the energy can be in fact injected into the system in a
radiative, thermal, or kinetic mode, and/or at different times
during the evolution of the system, and/or under different conditions.
These models are, nevertheless, similar in the aims of
removing gas from the galaxy, quenching star formation, and turning the
galaxy from blue to ``red and dead''.
AGN feedback has also been proven to predict the correct slopes
and normalizations of the observed scaling relations between
BHs mass and galaxy
properties \citep[e.g.,][]{SilkRees}, such as the stellar dispersion velocity $\sigma$ or
$K$-band luminosity, although feedback is \emph{not} a necessary condition
to substantiate such scaling relations \citep[e.g.,][]{MiraldaKoll}.


The \emph{physics} of accretion onto the central BH remains,
however, poorly understood. The problem of how to remove angular
momentum from gas in the inner regions of galaxies allowing it to
accrete onto the central BH, is still far from settled. Furthermore,
\citet{Thompson05} pointed out that even if gas is efficiently
transferred to the central regions through some global torque, then
the high densities will lead all gas to be easily converted into
stars at the 1--10 pc scale, leaving no fuel for the BH.


Most SAMs now include some basic recipes to account for BH growth
and the co-evolution with their host galaxies, although
the assumptions and results sometimes differ significantly among
different groups. Some
models predict, for example, a merger dominated growth of BHs
at late times, while others state that gas accretion always
dominates the evolution. \citet{Malbon07}, within the context
of their hierarchical model of galaxy evolution, find
that while lower mass BHs mainly grow through gas accretion during starbursts, more massive
BHs grow first by accretion at higher redshifts, and then predominantly by merging of pre-existing BHs at lower redshifts. This conclusion is not shared by other SAMs, which instead
grow all BHs mainly through gas accretion (e.g., \citealt{Lapi06}).

SAMs have adopted several different
prescriptions for triggering accretion onto the central BHs,
such as radiation-drag \citep[e.g.,][]{Kawakatu03},
bar-instabilities \citep[e.g.,][]{BBR84}, cloud-cloud
collisions \citep[e.g.,][]{Hop06fuel}, merging or galaxy
interactions \citep[e.g.,][]{Cav00,DiMatteo05,Croton06,Hop06fuel,Hopkins06}, surface density instabilities \citep[e.g.,][]{Cen07}, stellar
feeding \citep[e.g.,][]{MiraldaKoll,CiottiOstriker07}.
Nevertheless,
most SAMs provide reasonable matches to the \emph{same} observables, such
as the optical quasar luminosity function, although sometimes their input
BH physics might differ significantly.

While independent, fundamental tests for BH evolution will probably need to wait for future gravitational wave experiments, such as LISA (the ``Laser Interferometer Space Antenna'', a joint ESA/NASA mission), important
constraints can be set on BH evolution in ways independent of specific models, via basic accretion and merging models which also include clustering, as already extensively discussed in the previous sections.
In this section we instead briefly review some of the main
results achieved by several SAMs which attempted
to reproduce the overall BH evolution.

%


\subsection{UNDERSTANDING THE EVOLUTION AND SHAPE OF THE AGN LUMINOSITY FUNCTION}
\label{subsec|understandingtheAGNLF}

The two power-law observed AGN bolometric LF \citep[e.g.,][and references therein]{Hopkins07LF,SWM} is an
essential tool to probe BH evolution. In the overall, a significant
decline in the AGN luminous density is observed, with the faint-end
rapidly steepening at redshifts $z\lesssim 1.2$.

Such an evolution could be described by two extreme scenarios.
On the one hand, less massive BHs form continuously over time, to track the
population decline to lower luminosities while retaining nearly
constant (or slightly declining) Eddington ratios $\lambda\equiv
L/M_{\bullet}$, as in \citet{Kollmeier06}. On the other hand,
the strong AGN luminosity density redshift evolution could also be
explained by a distribution of already massive BHs shining with ever
lower luminosities over time, inevitably causing a significant
decrease in the overall Eddington ratio distribution. From the
simultaneous match to the quasar clustering and luminosity function,
\citet{WL05} concluded that bright quasars at $z\lesssim 2$
have a lower abundance than expected from the number of merging
halos, and discussed how a viable solution would be for quasars to
shine at sub-Eddington luminosities. Similar questions arise
regarding the steepening of the faint-end of the AGN LF: is it
produced by lower-mass BHs shining at the Eddington limit or is it
mostly due to high-mass BHs radiating at sub-Eddington luminosities?

Within the framework of their numerically-predicted BH light curves,
\citet{Hopkins06LF} found that the bright-end slope and break
luminosity time behavior in the AGN luminosity function, directly
reflects the formation rates of BHs shining close to their peak
emission following a major merger, while the main contribution to
the faint-end is from BHs shining at luminosities far below the
peak. This scenario would imply a continuous formation of massive
BHs down to very low redshifts driven by \emph{major} mergers. On
other grounds, however, extending major mergers as main triggers for
the overall quasar population at all times, might give rise to several
problems \citep[e.g.,][]{Bournaud07,Naab07}. 

\subsection{THE HIGH-REDSHIFT EVOLUTION: FAST GROWTH OF MASSIVE
GALAXIES AND BLACK HOLES} \label{subsec|highz}

At high-$z$, the main SAMs fail to grow BHs fast and massive enough
to reproduce the relatively high number of luminous quasars observed
at high-$z$ and their extreme clustering \citet{Shen07}.
High-resolution numerical simulations have shown that DM halos grow
in a two-phase mode \citep[e.g.,][]{Zhao03}. The inner core and
central potential of the host halo is built during a fast, chaotic,
merging phase of multiple subhalos after which a longer, smoother
accretion phase occurs. During the latter phase, mass is
preferentially added to the outskirts of the halo without affecting
the inner regions much. This behavior may help in preserving the
scaling relations between the central BH and host galaxy, and, in
general, limit the role of environment in the formation of luminous
galaxies \citep[e.g.,][]{Bournaud07,Tinker08,Cook09}. In the DM
potential wells, SAMs attempt to speed up the formation of massive
galaxies and their BHs by allowing baryons to cool in high clumping
factors \citep[e.g.,][]{Granato04,Maller04,Scannapieco05} and/or cold streams flowing directly to the
center \citep[e.g.,][]{DekelBirnboim}. For example, by using
enhanced clumping factors and short formation timescales, Granato et
al. (2004), \citet{Ciras05}, Lapi et al. (2006) and Shankar et
al. (2006), were able to reproduce several dynamical, photometric,
and statistical properties of BHs, spheroidal galaxies and AGNs.

At the same time, SAMs need to cope with the difficulty of
reproducing the non-linear metallicity behaviors observed in
early-type galaxies, with the more massive
galaxies being $[\alpha/Fe]$-enhanced with respect to the less
massive ones. The $Fe$ production is roughly proportional to the
(delayed) Supernova Ia rate. Thus, reducing the burst timescale in
the more massive galaxies, proportionally reduces their amount of
$Fe$ inducing an $[\alpha/Fe]$-enhancement (e.g., \citealt{Matteucci94}).
In practice, reproducing, and physically explaining, this
differential trend in metallicity is a challenge for SAMs. Granato
et al. (2004), similarly to other models such as those of \citet{Croton06}, De Lucia et al. (2006), \citet{Pipino09}, naturally
find the more massive galaxies to be, on average, the oldest and
most metal rich.

However, to get the right $[\alpha/Fe]$-enhancement some
extreme assumptions need to be made. Granato et al. (2004), for
example, allowed for Super-Eddington accretion at high redshifts
(this assumption has also been used in the recent numerical
simulations by \citealt{DiMatteo08}), thus boosting BH growth which
more rapidly meet the condition for self-regulation and quench star formation earlier than less massive galaxies. \citet{Granato06}
successfully adopted the same prescriptions to reproduce the bulk of
the $850\mu$ counts (see also \citealt{FontanotDust}) observed
with the submillimeter array \emph{SCUBA}, a difficult task for
hierarchical galaxy formation models \citep[e.g.,][]{Granato00}. By
allowing for substantial BH growth during the dust-obscured phase
triggered by radiation drag (\citealt{Kawakatu03}), they also
predicted the correct number counts of AGN sources emitting above a
certain X-ray flux limit, as measured by \citet{Alexander05}. Lapi
et al. (2006) using the same model to grow BHs to at $z\gtrsim 2$,
and a significantly large scatter in the quasar luminosity-halo
relation, were then able to reproduce the bright end of the observed
AGN luminosity function (see the previous Authors for details). However,
a large scatter in the luminosity-halo relation of high-$z$ quasars
might be difficult to reconcile with the extreme clustering signal
measured for those quasars (see \S~\ref{sec|clustering}), and more
secure data and more detailed modeling is required
to set final conclusions.

Similar conclusions on the growth of the most massive BHs at
high-$z$ were reached by Marulli et al. (2008; similar results were
found by Fontanot et al. 2006 using the MORGANA model). Although
their semi-analytical framework, developed on the outputs of the
Millennium Simulation, better tracks, in principle, DM merger
histories than the analytical treatments by Lapi et al. (2006),
their final conclusions on the baryonic physics is similar. In fact,
while they can reproduce the observed AGN luminosity function at low
and intermediate redshifts, at $z > 1$ they under-predict the number
density of bright AGN, regardless of the BH mass accretion rate and
light curve model assumed for each quasar episode. Despite the
considerable freedom in choosing the Eddington ratio distributions,
they failed to find a model able to simultaneously match the
observed BH scaling relations, the BH mass function, and the AGN
luminosity function, especially at high redshifts. Also allowing for
larger seed BH masses, they were still not able to fit the high-$z$
luminosity function. They interpreted this failure as an indication
that the present theoretical framework is itself inadequate to
account for the full AGN phenomenon. Analogously to Lapi et al. and
Granato et al., \citet{Marulli08} also suggested that a
significant improvement in the model might arise by assuming an
accretion efficiency that increases with redshift, for example by
forming more concentrated disks in the past, making more efficient
the BH feeding (see, e.g., \citealt{Mo98}).

\subsection{THE INTERMEDIATE REDSHIFT EVOLUTION:
INTERACTIONS, BLACK HOLES RE-ACTIVATIONS, ACTIVITY IN DISKS}
\label{subsec|interactions}

Although efficient, the fast accretion phase occurring during the
epoch of rapid merger events in the high-redshift Universe cannot be
the only viable formation mode to fully grow all local BHs.
For $z \lesssim 1.5$ major mergers become much
rarer at galactic scales; fewer new massive BHs are then expected to
be formed. The prevailing dynamical events that could still trigger
star formation in a galaxy and possible gas accretion onto the
central BH, are best described as \emph{interactions} between
galaxies, occurring in the small, dense groups that begin to form at
these epochs. The accretion is also expected to be supply-limited,
due to progressive dilution of the host gas reservoir and lack of
fresh gas resupply from gas-rich
mergers. 

As discussed in \citep[e.g.,][and references therein]{Cav00,Hop08clust}, small galactic groups with
virial mass $\gtrsim 10^{13}\, {\rm M_{\odot}}$ could provide
particularly suitable sites for the hosts to interact with their
companions. An interaction of the host with a group companion will
perturb the galactic potential, and destabilize a fraction $f$ of
the cold gas mass $m(z)$ in the host from kpc scales to the inner
regions, possibly triggering a circumnuclear starburst (see, e.g.,
the numerical simulations by \citealt{Mihos99}, and the IR observations by
\citealt{Franceschini93}) and reactivating the central BH to a level
of luminosity $L\propto fm(z)$. Direct evidences of star-formation
activity at intermediate redshifts connected with clearly interacting galaxies have now been
provided by several groups \citep[e.g.,][and references therein]{Urrutia08}, who also find that a large fraction of AGNs have close companions or show signs of
ongoing or recent interactions, although locally and/or at lower luminosities
this association between AGN activity and close companions might break down, as discussed in \S~\ref{sec|clustering}. 

\citet{Vittorini05} have carried out a detailed semi-analytical
computation of interactions and minor/major merging of galaxies
within the framework of the hierarchical growth of dark matter
halos. They found that the strong luminosity decline with redshift
of the AGN luminosity function, given by
$L\propto f\times m(z)$, is fully accounted for by the continuous
exhaustion of gas $m(z)$ driven by galaxy interactions. The highly
\emph{stochastic} nature of the accretion events described by
interactions, parameterized in the factor $f$, predicts a largely
increasing scatter in the Eddington ratio distributions at lower
redshifts, peaked at lower
and lower accretion rates, in line with the several recent empirical
findings discussed in \S~\ref{subsec|BHMFevol}.

Other possible ways to reproduce the bright end of the quasar
luminosity function have been extensively discussed in, e.g., \citet{Scannapieco04}, and \citet{Croton06}. These
models mimic the fall-off and steepening of the quasar luminosity
function at low redshifts by assuming that late AGN feedback in massive halos prevents cooling flows in massive halos (the so-called ``radio-mode'' feedback).
More recently, \citet{Cav07} have presented the full predictions of their semi-analytic
model for the co-evolution of BHs and galaxies, which includes
galaxy interactions, among other dynamical triggers. Their model
predicts a ``bimodal'' Eddington ratio distribution, with a peak
around $\lambda\sim 0.01$, mainly contributed by low-luminosity AGNs
in red galaxies, and a peak at $\lambda\lesssim 1$ contributed by
lower mass BHs in bluer, star-forming galaxies.

It is clear, especially after the significant results obtained
from SDSS data \citep[e.g.,][]{Kauffmann03,Kauffmann09,Li06,Li08}, that a significant part of the AGN activity
in the local Universe must be triggered by
other mechanisms other than major mergers, such as
bar instabilities, radiation drag, stellar winds.
Moreover, observations have unveiled the presence of AGNs also in isolated disk-dominated
galaxies \citep[e.g.,][]{Satyapal08}, as discussed in \S~\ref{subsec|BHsInLaterTypeGalx}.
AGN feedback might have been
possibly efficient due to the lower mass in the central BHs
\citep[e.g.,][]{Kawakatu04,Shankar06}, and the metal-rich gas outflow may be
expected to recollapse towards the center on a relatively short
timescale, due to the shallower potential wells that characterize the
halos hosting disk galaxies today.
\citet{Tsujimoto07} finds that if one includes a double infall of
primordial and enriched gas in chemical evolution models of the
Galaxy, with the enriched outflow occurring after 1 Gyr since the
onset of the star formation, it is possible to explain the observed
$\alpha$-element abundances and their non-linear behavior in the
Galaxy as a function of iron abundance. This model can
also simultaneously reproduce the sharply peaked metal abundance of disk
stars against iron abundance.

To sum up, most models agree in associating the luminous, high-$z$ quasars
with fast, gas-rich merging events, which rapidly built up the host potential
wells and the scaling relations discussed in \S~\ref{sec|correlations}.
A longer and milder activity phase has
probably occurred at lower redshifts, during which in-situ and milder dynamical events
have regulated the late evolution in the AGN/BH population. However, as discussed
in this section, many details of this basic picture are still missing,
and are currently the subject of intense investigation.

\section{Conclusions}
\label{sec|conclu}

This Review addresses the main topics and aims concerning the
evolution of the BH population, and discusses several ways to
constrain it in ways independent of specific models of galaxy evolution.
The main results can be listed as follows.
\begin{itemize}
\item The local BHMF can be now rather accurately
assessed. In more detail, the BHMF hosted in early-type
galaxies can be obtained on exploiting the velocity dispersion or
luminosity functions of host galaxies, coupled with the
$M_{\rm BH}$--$\sigma$ or $M_{\rm BH}$--$L_{\rm sph}$
relationships, respectively, yielding similar results
within $\sim 30\%$ accuracy. The
contribution from BHs hosted by late-type galaxies is more
uncertain, and is mostly confined to the low mass end of the BHMF. The
overall BH mass density amounts to $\rho _{\rm BH}=(3-5)\times 10^5 M_{\odot}/\hbox{Mpc}^3$ (\S~\ref{sec|BHMF}).
\item The match both between the local BH mass density and the one predicted by integrating the bolometri AGN
luminosity function (Soltan 1982) yields interesting
constraints for a mass to radiation conversion efficiency
$\epsilon \lesssim 0.1$,
although systematic uncertainties on bolometric corrections, luminosity functions, and the local BHMF prevent more precise estimates (\S~\ref{sec|evolution}).
\item The average accretion history of BHs tracks the average SFR, with a
proportionality
constant of the order of the local one. Several works have
empirically proven that the ratio between the degree
of evolution in the local relations is relatively
mild, suggesting that the bulk of the BH and host galaxy populations should
have mostly co-evolved, at least up to $z\lesssim 2$ (\S~\ref{sec|evolution}).
\item The \emph{shape} of the local BHMF yields additional interesting constraints
on the typical Eddington ratio $\lambda\sim 0.5$.
The alternative possibility that most of the mass locked in BHs
in the local Universe has been
accumulated by ``dark'' accretion (i.e., accretion undetectable by
either optical or hard X-ray surveys, as in the case of BH
coalescence), is severely constrained by the above results. In order
to make room for this possibility one has to assume that the
radiative efficiency during the visible AGN phases is at the
theoretical maximum of $\epsilon\simeq 0.3$--0.4. But even in this
case (unless the bolometric correction is far lower than currently
estimated) the contribution of radiative accretion to the local BH
mass density is $\ge 25\%$, and one is left with the problem of fine
tuning the radiative and non-radiative contributions in order not to
break down the match with the local BHMF obtained with radiative
accretion alone. One would also face the problem of accounting for
the tight relationships between BH mass and mass or velocity
dispersion of the spheroidal host, naturally explained by feedback
associated to radiative accretion. 
The analysis of the accretion history highlights that the most
massive BHs (associated to bright optical quasars) accreted their mass
faster and at higher redshifts (typically at $z>1.5$), while the
lower mass BHs have mostly grown at $z<1.5$ (\S\S~\ref{subsec|BHMFevol} and \ref{subsec|merging}).
\item Quasar clustering provides additional, independent constraints on
the main parameters governing the BH evolution. It has been shown how AGN clustering can set important constraint on the
possible redshift and/or mass dependence of the typical \lam\ and $\epsilon$, thus helping us in discriminating among a number of successful models. Clustering is in fact a
direct measure of the masses, and therefore number densities, of the
halos hosting the quasars. Matching the quasar host number density
with the one derived from accretion models (Eq.~[\ref{eq|conteq}]), in turn yields
interesting constraints on radiative efficiency and Eddington ratios
as a function of time and/or BH mass (\S~\ref{sec|clustering}).
\item Most galaxy formation models now include some basic recipes to account for BH growth
and the coevolution with their host galaxies, although it has been
discussed in this Review that
the assumptions and results sometimes differ significantly among
different groups (\S~\ref{sec|models}). The model-independent results discussed above will
help shedding light on the true physical mechanisms which
regulate BH growth across time.

\end{itemize}

\section*{Acknowledgments}
I would like to warmly thank here many collaborators whom I had the pleasure to work with over
the years, Mariangela Bernardi, Silvia Bonoli, Alfonso Cavaliere, Martin Crocce, Xinyu Dai, Luigi Danese, Gianfranco De Zotti, Laura Ferrarese, Gianluigi Granato, Zoltan Haiman, Andrea Lapi, Cheng Li, Federico Marulli, Smita Mathur, Jordi Miralda-Escud\'{e}, Jorge Moreno, Paolo Salucci, Yue Shen, Ravi K. Sheth, Gregory R. Sivakoff, Valerio Vittorini, David H. Weinberg.
I acknowledge the Alexander von Humboldt Foundation and NASA Grant NNG05GH77G for support. I finally thank Lucas Bakker for assistance during the editing of the manuscript.

\bibliographystyle{elsarticle-harv}
\bibliography{RefMajor}

\end{document}